\documentclass[onecolumn]{emulateapj}

\def\persqcm{$\rm cm^{-2}$}

\def\h2{$\rm H_2$}
\def\error{$\pm$}
\def\e#1{$\times 10^{#1}$}

\def\asec{\arcsec}

\def\deg{\arcdeg}
\def\thr{$^h$}
\def\tmin{$^m$}
\def\tsec{$^s$}

\def\kms{km~s$^{-1}$}
\def\persec{s$^{-1}$}
\def\peryr{yr$^{-1}$}
\def\solmass{$\rm M_{\sun}$}

\newcommand{\htooco}{$\rm H_2/CO$}
\newcommand{\htoo}{$\rm H_2$}
\newcommand{\convunits}{$\rm cm^{-2}\,(K\,km\,s^{-1})^{-1}$}

\newcommand{\jbkms}{$\rm Jy\,b^{-1}\,km\,s^{-1}$}
\newcommand{\jykms}{$\rm Jy\,km\,s^{-1}$}
\newcommand{\mjb}{$\rm mJy\,b^{-1}$}

\newcommand{\msunsqpc}{$\rm M_\odot\:pc^{-2}$}
\newcommand{\halpha}{H$\alpha$}
\newcommand{\hii}{H{\sc II}}
\newcommand{\SigHI}{$\Sigma_{\mathrm{HI}}$}
\newcommand{\SigHtoo}{$\Sigma_{\mathrm{H2}}$}

\begin{document}

\title{The Evolution of the ISM in the Mildly Disturbed Spiral Galaxy NGC
4647}
\author{L. M. Young}
\affil{Physics Department, New Mexico Institute of Mining and Technology,
 Socorro, NM 87801}
\email{lyoung@physics.nmt.edu}
\author{E. Rosolowsky}
\affil{Harvard-Smithsonian Center for Astrophysics, 60 Garden St., MS-66,
Cambridge, MA 02138}
\email{erosolow@cfa.harvard.edu}
\author{J. H. van Gorkom}
\affil{Astronomy Department, Columbia University, 538 W. 120th St., New
York, NY 10027}
\email{jvangork@astro.columbia.edu}
\author{S. A. Lamb}
\affil{Physics Department, University of Illinois at
Urbana-Champaign, 1110 W. Green St., Urbana, IL 61801}
\email{slamb@astro.uiuc.edu}

\slugcomment{ApJ, Accepted 15Jun06}

\begin{abstract}  
We present matched-resolution maps of HI and CO emission in the Virgo
Cluster spiral NGC~4647.  
The galaxy shows a mild kinematic disturbance in which one side of the
rotation curve flattens but the other side continues to rise.
This kinematic asymmetry is coupled with a dramatic
asymmetry in the molecular gas distribution but not in the atomic gas.
An analysis of the gas column densities and the interstellar pressure 
suggests that the \htoo/HI surface density ratio
on the east side of the galaxy is three times higher than expected from the hydrostatic
pressure contributed by the mass of the stellar disk.
We discuss the probable effects of ram pressure, gravitational
interactions, and asymmetric potentials on the interstellar medium
and suggest it is
likely that a $m=1$ perturbation in the gravitational potential could be
responsible for all of the galaxy's features.  
Kinematic disturbances of the type seen here are common, but the
curious thing about NGC 4647 is that the molecular distribution appears
more disturbed than the HI distribution.  
Thus it is the combination of the
two gas phases that provides such interesting insight into the 
galaxy's history and into models of the interstellar medium.

\end{abstract}

\keywords{
galaxies: individual (NGC 4647) ---
galaxies: spiral ---
galaxies: ISM ---
galaxies: kinematics and dynamics ---
galaxies: evolution ---
ISM: molecules
}

\section{Introduction}

Galaxies that are rich in cold gas tend to have
roughly equal amounts of gas in the atomic and in the molecular phase.
But, to the best of our knowledge, star formation only occurs in the
molecular phase.  Therefore, the global balance between atomic and
molecular
gas is crucial to the long-term evolution of galaxies and their
stellar populations.  What determines the balance between
atomic
and molecular gas in a galaxy, and how does that balance evolve as the
environment of the galaxy changes?

On theoretical grounds we would expect that the density and temperature of the
gas, the strength of the dissociating UV field, and the metallicity all
affect the relative amounts of molecular and atomic gas
\citep{hidaka2002,elmegreen93}.  
The gas should also be in hydrostatic equilibrium
with a pressure that is mostly determined by the stellar distribution,
since stars contribute the bulk of the mass in the inner parts of galaxies
where the \htoo---HI transition occurs.

If a galaxy is disturbed by gravitational interactions or by
falling into an intracluster medium, 
the balance between atomic and molecular gas may be altered.  
Indeed,
\citet{miller03a} and \citet{miller03b} observe increased star formation
activity in the members of a galaxy group or subcluster that recently
merged into a larger cluster.
In a
bottom-up (cold dark matter) cosmology such disturbances should
be common, and they may be important both to the star formation histories
and the morphological evolution of galaxies.
In turn, the disturbances and asymmetries in galaxies can inform our
understanding of cosmology as they allow us to place constraints on the
merger/interaction rate.

The semi-analytic simulations of galaxy formation
(e.g.\ \citet{somerville99,kauffmann93} and successors)
are attempts to understand the observed properties of galaxies by
predicting their star formation histories in a cosmological context.
To this end the semi-analytic simulations typically
assume that the star formation rate can be calculated from
empirically motivated recipes involving the gas content and the recent
merger history.  However,
these empirical recipes may not capture important phases of
galaxy evolution. Thus it is desirable to have a more physically motivated
understanding of the interstellar medium (ISM) and star formation that can
be applied to both semi-empirical studies and future detailed
N-body/hydrodynamic simulations.

\citet{krumholz05} have recently put together a detailed model for the star formation
rate in galaxies. 
They begin with the molecular cloud microphysics, 
in which fractal turbulence drives
a small fraction of the gas to high enough density that it
collapses to form stars.
They extend the physics to galaxy-wide scales by then considering the
molecular clouds as
gravitationally bound entities formed in a disk that is marginally 
stable against gravitational collapse using the Toomre criterion 
\citep{Toomre64,BT}.
We note that this model still depends critically on knowing how much of the
gas disk is in molecular form and how much is atomic.  In the 
absence of observations, that information must be obtained from a
hypothesis such as that of \citet{br04}, 
which predicts atomic and molecular column density ratios as a function of
the midplane hydrostatic pressure in a disk.
Future work on simulations of galaxy evolution could be made more realistic
by incorporation of such
physically-based models of a multiple-phase ISM and its effects on the
star formation rate.

Our current contribution to this effort is an observational study
that begins to test such physical models on a
gently disturbed galaxy.
We present
an analysis of the kinematics of the neutral interstellar
medium in the Virgo Cluster spiral NGC 4647, along with an analysis of the
relationships between atomic gas, molecular gas, and star formation in the galaxy.  
We investigate whether the observed molecular/atomic balance and the
disturbed kinematics can be
understood in the context of the galaxy's environment.
The form and magnitude of the kinematic disturbance in NGC 4647 are 
common, so the
processes affecting this galaxy might be applicable to many spirals.
However, few mildly disturbed spiral galaxies have been
mapped at matching resolutions in both atomic and molecular gas, 
so opportunities for this type of
analysis are rare.

\section{Meet NGC 4647}

\begin{figure}
\epsscale{0.8}
\plotone{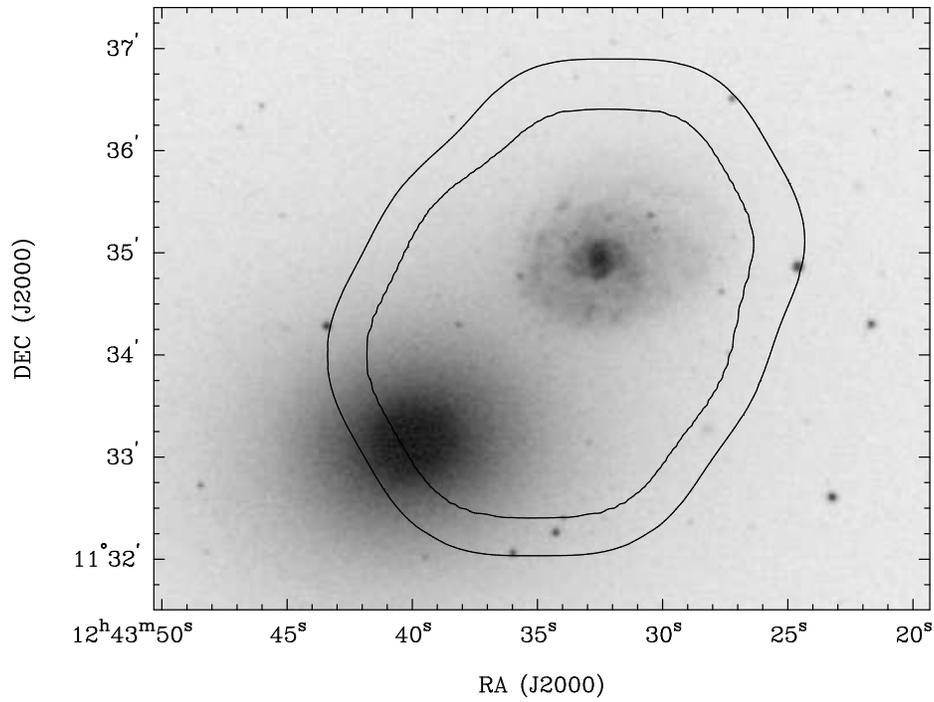}
\caption{Mosaiced field of view for the CO data.  The nominal sensitivity to a point
source (the primary beam, or the gain of the image) is 0.99 within the
inner contour and falls to 0.5 at the outer contour.  Contours are
superposed on an image of the system from the blue plates of the
Digitized Sky Survey.  The primary beam of the VLA at 20 cm is much
larger; the sensitivity of the HI data falls to 0.5 at a distance
of 15.5\arcmin\ from the pointing center and it is effectively uniform over
the region shown here. 
\label{fieldofview}
}
\end{figure}

NGC 4647 is a gas-rich Virgo Cluster spiral whose interstellar medium has
been gently disturbed and whose surroundings suggest some possible causes of
the disturbance.  
It is 3.2\deg\ = 0.95 Mpc in projection from M87.
The galaxy's distance has been estimated via the Tully-Fisher method to be 
$17.3^{+3.0}_{-2.6}$ Mpc \citep{gavazzi99}, so that it is close to
the midplane of the Virgo Cluster.
It is also notable for being in a close pair
with the elliptical galaxy NGC 4649 (M60); the two are 2.6\arcmin\ apart
or 13 kpc in projection (Figure \ref{fieldofview}).
The galaxies are close in radial velocity as well, with NGC~4647 at 
1415 \error 4 \kms\ and NGC~4649 at 1114 \error 17 \kms\
\citep{RC3}.
Surface brightness fluctuations give the distance to NGC 4649 to be 
16.8 \error 1.2 (random) \error 1.2 (systematic) Mpc \citep{tonry01}, so this
independent distance measurement is also consistent with the two galaxies
being physically close to each other.
We therefore adopt a distance of 17 Mpc to both galaxies and entertain the
possibility that they are gravitationally interacting.

The location of NGC 4647 in the midst of the Virgo Cluster also suggests
that the galaxy may have suffered ram pressure stripping by the
intracluster medium.
The galaxy is mildly HI-deficient; according to \citet{kenney89}
the HI mass is a factor of 3 lower than what one would expect for
an isolated galaxy of the same optical diameter.  
This deficiency is very typical 
of late-type spirals that are 3\deg\ in projection from M87.
The molecular/atomic mass ratio is also typical for galaxies of the same HI
deficiency \citep{kenney89}. 

The stellar distribution of NGC 4647 (traced by an R-band image) and \halpha\ emission
are both modestly asymmetric, with an enhancement in \halpha\ surface
brightness at the outer edge of the star-forming disk on the side closer to
NGC 4649 \citep{koopmann2004}.
Here we look in greater detail at the gas distribution and kinematics to
search for clues to the processes responsible for disturbing the ISM in NGC
4647.

\section{Observations and Data Reduction}

\subsection{CO imaging}

$^{12}$CO 1-0 emission from the NGC 4647--4649 system was imaged 
with the 10-element Berkeley-Illinois-Maryland
Association (BIMA) millimeter interferometer at Hat Creek, CA
\citep{welch96}.
The BIMA observations were carried out in the D configuration (projected 
baselines 2.3 to 11 k$\lambda$) for a total of 5.5 hours on 2001 June 15,
4.25 hours on 2003 June 8, and 2 hours on 2003 June 28.
Observations were made in mosaic mode using a hexagonal grid of 7
pointings with center-to-center separations of 60\arcsec.
The 2001 observations were made with a pointing center of 
12\thr 43\tmin 35.0\tsec, +11\deg 34\arcmin 00\asec (J2000),
and the 2003 observations were made with a pointing center of
12\thr 43\tmin 32.0\tsec, +11\deg 35\arcmin 07\asec.
This coverage was designed to give sensitivity to CO emission
throughout the disk of NGC 4647 as well as in the region between the
two galaxies and the central regions of
NGC 4649.  \citet{sage89} reported a single-dish CO detection of
the elliptical NGC 4649, so the mosaic observations were partly designed to
investigate those claims. 
Figure \ref{fieldofview} shows the field of view of these data.
The velocity range covered by the CO data is 888 to 1897 \kms.

Reduction of the BIMA data was carried out using standard tasks in the
MIRIAD package \citep{sault95}.
Absolute flux calibration was obtained from observations of Saturn and the
secondary flux calibrator 3C273.
Phase drifts as a function of time were corrected by means of a
nearby calibrator observed every 30 to 40 minutes. 
Gain variations as a function of frequency were corrected by the online
passband calibration system;
inspection of the data for 3C273 indicate that residual
passband variations are on the order of 10\% or less in amplitude and 2\deg\ in
phase across the entire band.

The calibrated visibility data were weighted by the inverse square of the
system temperature, then Fourier transformed.
The MIRIAD task {\it invert} performs an appropriately weighted linear
combination of the data from the different pointings.
No continuum emission was evident in the line-free channels, so no attempt
at continuum subtraction was made.
The dirty images were deconvolved using both the Clark clean algorithm and
a maximum entropy algorithm, but the two methods gave indistinguishable
results.
The resulting image cube's resolutions and noise levels are given in Table
\ref{obstable}. 
Integrated intensity images were produced by the masking 
method: 
the deconvolved image cube was smoothed along both spatial and velocity 
axes, and the smoothed cube was clipped at about 2.5$\sigma$ in absolute
value.
The clipped version of the cube was used as a mask to define a
three-dimensional volume in which the emission is integrated over velocity.
This masking method for producing moment maps
is described in greater detail by 
\citet{wong01} and by \citet{regan01}.
Integrating the ``moment zero" image again over the spatial directions then gives the total CO flux
of the galaxy.
In addition, a velocity field is constructed by fitting a Gaussian line
profile to the CO spectrum at each position within the galaxy.

The total CO flux we measure from NGC 4647 is 
450 \jykms\ with an uncertainty of
10--15\% from absolute calibration; it is consistent with the
previous single-dish measurement.  \citet{kenney88} used the
FCRAO telescope, with a beam FWHM of 45\asec, and they
quote a total flux of 600\error 120 \jykms.  Within the error bars,
these two measurements agree.  It is important to note,
however, that the flux quoted by \citet{kenney88} has been multiplied up by
a factor of 1.5 to account for ``unobserved'' gas.  This correction
factor is an educated guess of how much emission was
not sampled by their 5 pointings (spaced by the FWHM) 
along the galaxy's optical major axis.
We can make a more accurate comparison of the fluxes 
by convolving our integrated intensity map with a 45\asec\ Gaussian,
mimicking the sensitivity pattern of the FCRAO telescope.  When this is
done, we predict that the FCRAO telescope should measure 212 \jykms\
from the central position on the galaxy.  This estimate is in excellent 
agreement with the measured value of 219 \error 35 \jykms.
We conclude that the BIMA interferometer has not missed any substantial
component of CO flux from NGC 4647.
For 17 Mpc distance and 
a CO-to-H$_2$ conversion factor of 2.0\e{20} molecules cm$^{-2}$
(K~\kms)$^{-1}$ the \htoo\ mass is then $(1.0 \pm 0.15) \times 10^9$
\solmass.  The mass of helium can be included by multiplying up by a factor
of 1.36.

\subsection{HI imaging}\label{HIobs}

HI emission from NGC 4647 was observed with the 
National Radio Astronomy Observatory's Very Large Array (VLA)\footnote{The 
National Radio Astronomy Observatory
is operated by Associated Universities, Inc., under cooperative agreement with
the National Science Foundation.} in its D and C configurations on 2002
January 4 and 2002 December 2 and 7.
We obtained 3.7 hours on source in the D configuration and 7.2 hours on
source in the C configuration.
The system was observed with one pointing centered in between the two
galaxies.
The initial D configuration observations were made with a wide bandwidth, 6.25 MHz,
in order to cover the velocities of both NGC 4647 and NGC 4649; imaging of these data revealed 
that all of the HI emission from the system was coming from NGC 4647.
Subsequent C configuration observations were made with
a smaller 3.125 MHz bandwidth (centered on the velocity of NGC 4647)
and higher velocity resolution.
A nearby point source was observed every 30 to 45 minutes as a phase calibrator.
The absolute flux scale was set by observations of J1331+305
and bandpass calibration was determined from the same source.
All data calibration and image formation was done using standard calibration
tasks in the AIPS package.
Initial imaging revealed which channel ranges were free of HI line emission.
Continuum emission was subtracted directly from the raw uv-data by making first
order fits to the line-free channels.

The calibrated data were Fourier transformed using several different uvdata weighting
schemes chosen to enhance the spatial resolution or the sensitivity to large-scale
structures.  
A low resolution HI image was made covering
the entire 6.25 MHz bandwidth using the D configuration data alone and the
``natural" {\it uv} weighting scheme (all data points are assigned equal
weight).
A medium resolution HI image was made by combining all the C and D configuration
data with natural weighting.
The highest resolution HI image uses the C configuration data alone and
Briggs's robust weighting scheme \citep{SIRA2}, as implemented in the 
AIPS task IMAGR with a robustness parameter 0.3.
Final images' resolutions and sensitivities are given in Table
\ref{obstable}.  Dirty images were cleaned down to a residual level of 
approximately 1.0 times the rms noise fluctuations.
Integrated intensity (moment) maps and velocity fields were made in a similar manner as for the
CO images.

\begin{deluxetable*}{lcccc}
\tablewidth{0pt}
\tablecaption{Image Parameters
\label{obstable}}
\tablehead{
\colhead{} & \colhead{CO} & \colhead{HI: high res} & \colhead{HI: med res} & \colhead{HI: low res}  \\
}
\startdata
Velocity range, \kms & 898 -- 1897 & 1093 -- 1738 & 1093 -- 1738 & 626 -- 1914 \\
Velocity resolution, \kms & 10.09 & 10.4 & 20.8 &  20.8 \\
Beam size, \asec & 14.79 $\times$ 13.38 & 15.63 $\times$ 13.55 & 36.00
$\times$ 30.69 & 45.75 $\times$ 43.75 \\
Beam position angle, $^\circ$ & $-$14.3 & $-7.8$ & $-2.5$ & $-16.9$ \\
Linear resolution, kpc & 1.2 $\times$ 1.1 & 1.3 $\times$ 1.1 & 3.0 $\times$
    2.5 & 3.8 $\times$ 3.6 \\
rms noise, \mjb & 100 & 0.35 & 0.18 & 0.43 \\
N(HI) sensitivity, cm$^{-2}$ & \nodata & 5.7\e{19} & 1.1\e{19} & 1.5\e{19} \\
N(H2) sensitivity, cm$^{-2}$  & 2.8\e{20} & \nodata & \nodata & \nodata \\

\enddata
\tablecomments{All velocities in this paper are in a heliocentric system
using the optical definition.
The distance to NGC 4647 is taken to be 17 Mpc.  Column density
sensitivities are 3$\sigma$ in one channel; the $H_2$ limit assumes
a CO-to-H$_2$ conversion factor of 2.0\e{20} molecules cm$^{-2}$ per K
\kms\ of brightness temperature in $^{12}$CO 1-0.}
\end{deluxetable*}

The highest sensitivity HI cube (medium resolution, combined C+D configurations)
contains a total of 7.0 \error 0.1 \jykms\ of HI emission from NGC 4647.
The uncertainty in this value comes from a spread among
different techniques for making 
the integrated intensity image and from different sum regions.
The lowest resolution HI cube gives an integrated flux of 6.9 \jykms,
consistent with the above. The highest resolution HI cube also gives
a flux consistent with the above
(7.1 \error 0.3 \jykms).  
Accounting for uncertainty in the absolute flux calibration,
we derive a total flux of 7.0 \error 0.2 \jykms\ of HI from NGC 4647.
At 17 Mpc, this flux corresponds to $(4.8\pm 0.1)\times 10^8$ \solmass\ of HI.

\begin{figure}
\epsscale{0.8}
\plotone{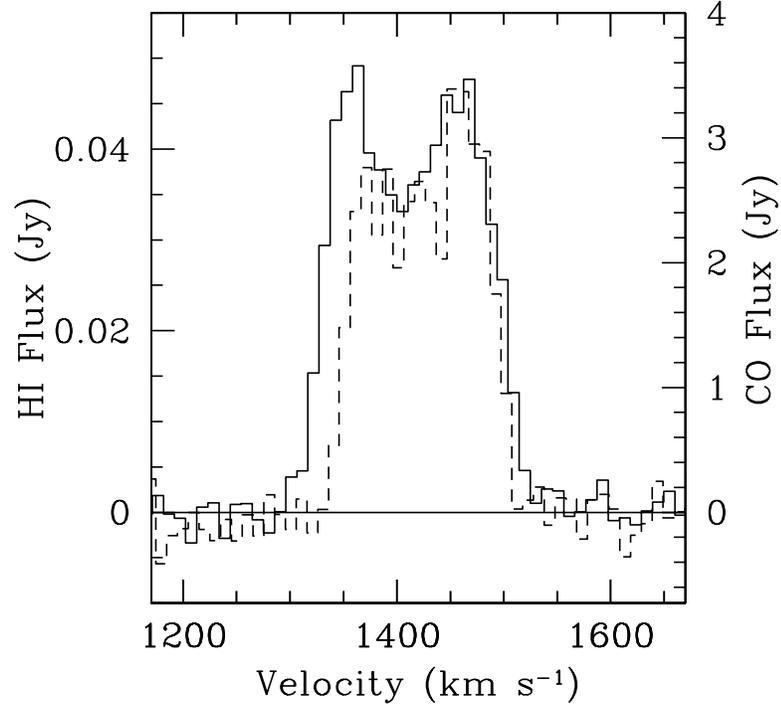}
\caption{HI and CO spectra of NGC 4647.  Atomic gas is indicated with the
solid line and the scale on the left side; molecular gas is indicated in the
dashed line and the right side scale.  Spectra are constructed by first
defining the spatial region of emission in an integrated intensity image and
summing over the same spatial region for all channels.
Velocities are heliocentric, in the optical definition.
\label{spectra}
}
\end{figure}

\citet{helou1984} measured the HI flux of NGC 4647 with the Arecibo telescope and 
found 8.2 \jykms\ with an error of 10\% to 20\% (not including
absolute calibration uncertainty).
\citet{warmels1988} used the WSRT to measure a total HI flux of 6.6 \jykms\ with an
error of 15\% to 20\%.
Earlier D configuration VLA observations of NGC 4647 are reported by
\citet{Cayatte90},
who claim an HI flux of 10.10 \error 0.44 \jykms.  It is not clear
why the 1983 VLA dataset would give such a different flux from the current
data,
but we note that our present observations have about 5 times
better sensitivity than those 1983 data when measured at the same spatial resolution.
Thus, we believe that our new VLA map has recovered all of the HI emission from
the galaxy; it is significant to note that even the highest
resolution HI cube still detects all of the HI emission.
Figure \ref{spectra} shows the integrated HI and CO spectra of NGC 4647.

\section{Results}

\subsection{HI distribution}\label{HIdist}

Figure \ref{tilefig} shows the HI column density  in NGC 4647 along with
images of the 
CO intensity, an R-band image, and continuum-subtracted H$\alpha$, all at
the same linear scale.
We have also computed the radial distribution of HI in NGC 4647 from
azimuthal averages over elliptical annuli in the approaching and receding halves of the galaxy
(Figure \ref{radialhi}).  ``Approaching" and ``receding" are defined with
respect to the adopted center of the galaxy and the kinematic major axis
that are described in Section \ref{kinematics} below.

The HI column density in NGC~4647 is relatively flat at 7.5 \msunsqpc\
in the interior of the galaxy.  (Gas surface densities quoted in \msunsqpc\
in this paper are corrected to face-on and include a factor of 1.36 for
helium; column densities in \persqcm\ are as observed.)
One notable feature of the HI distribution is an arm or segment of a
ring with enhanced column densities at radii of 30--45\asec\ = 2.5 to 3.7
kpc on the east side of the galaxy.  HI column densities are 10\% to 20\% higher in the arm than in the
central plateau; the highest HI column densities in the galaxy are here.
While the arm is strongest on the east and north sides of the galaxy, it is weak
or nonexistent on the western side.  
Beyond the arm, on the north and east sides of the
galaxy, the HI column density declines steeply; it falls below 
1.25\e{20} \persqcm\ at a radius of 67\asec\ (5.5 kpc), just a little over
one FWHM beyond the ridge of the arm.

The second notable feature of the HI distribution in NGC 4647 is a very
shallow dropoff in column density toward the southwest.
There is no well defined ridge or arm on this side, and whereas the HI
extends to 67\asec\ on the northeast side of the galaxy it peters out to
90\asec\ (7.4 kpc) on the southwest side.
Both radii are measured at the same 1.25\e{20} \persqcm\ in the data cube
with 15\asec\ resolution.
At low resolution (45\asec), the HI column density drops below the
sensitivity limit of $\sim 2\times 10^{19}$ \persqcm\ at 130\asec\ (10.7 kpc)
radius on the southwest side; however, there is relatively little HI at radii
beyond 90\asec\ since the low resolution map does not contain any more flux
than the high resolution map (Section \ref{HIobs}).
The larger HI extent on one side is reminiscent of the 
asymmetric HI in M101 \citep{baldwin80}.
Curiously, though, the southwest side of NGC 4647 has the shallowest HI
gradient but this is also the side with the steepest decline in CO intensity
(Section \ref{comorph}).

In the earlier map of NGC 4647 by \citet{Cayatte90},
the HI appeared to be slightly extended in the direction of
NGC 4649.  Those authors suggested that this extension could be caused by
gas transfer from NGC 4647 to NGC 4649,
possibly fueling the nuclear activity in NGC 4649.
Our present images show no sign of any HI extension towards NGC 4649, so we
conclude that we are not presently observing gas transfer from the spiral to the
elliptical.

\subsection{CO distribution}\label{comorph}

\begin{figure}
\epsscale{0.9}
\plotone{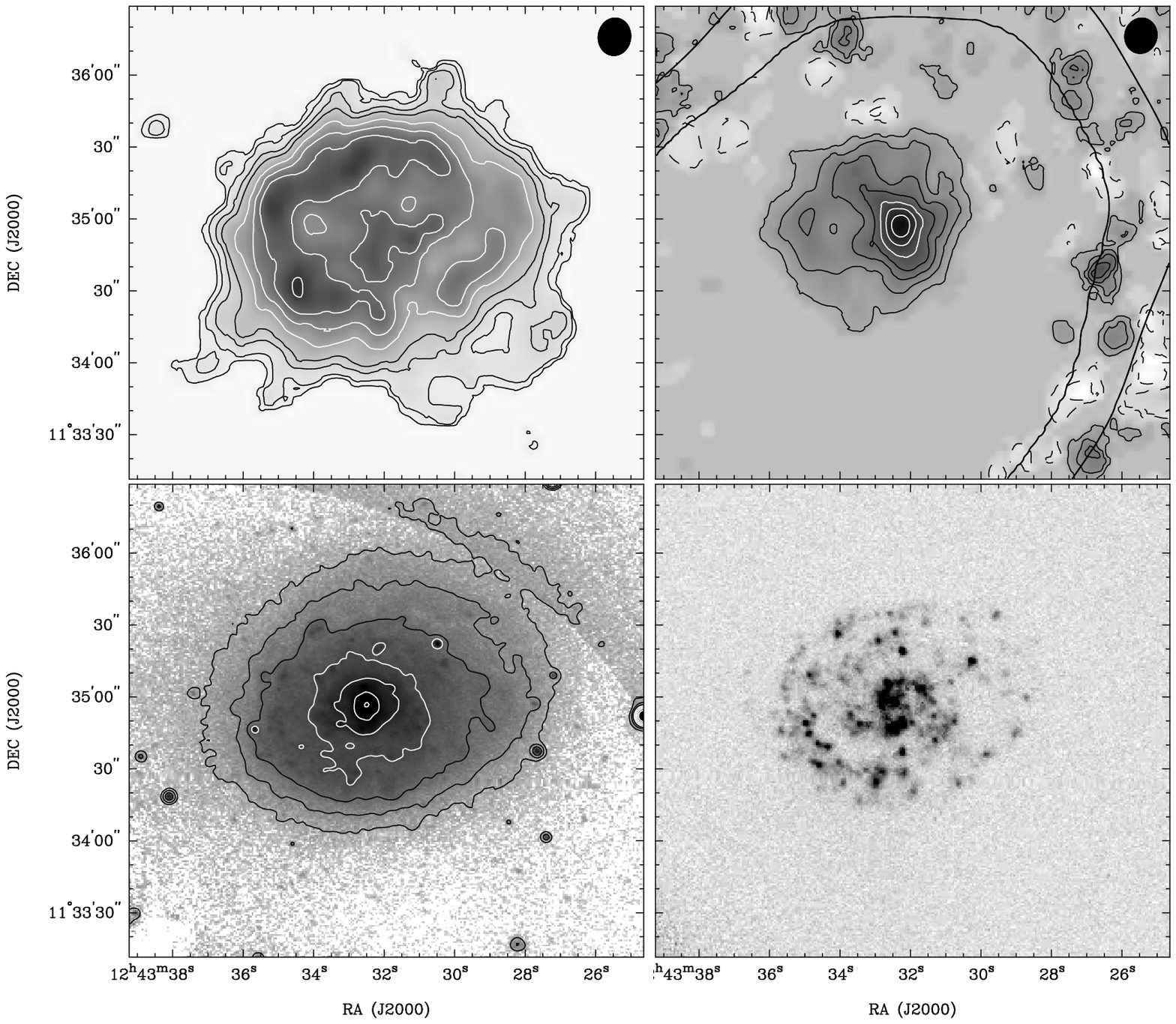}
\caption{HI, CO, stellar, and H$\alpha$ emission from NGC 4647.
Top right: HI column density from the high resolution (15\asec) data cube.  Contours are 0.5,1,2,3,5,7,9, and 12
\e{20} \persqcm.  
Top left: CO column density.  Assuming a \htooco\ conversion factor
of 2.0\e{20} \convunits, a CO intensity of 7.2 \jbkms\ is equivalent to
an \htoo\ column density of 6.7\e{20} \persqcm.
Contours are ($-2$, $-1$, 1, 2, 3, 4, 5, 6, 7)
$\times~6.7\times 10^{20}$ \persqcm.
The gain=99\% and 50\% contours are indicated as well with solid black
lines; note that the noise level increases dramatically in the regions
beyond gain=99\% because of large drops in the effective time-on-source at
those distant positions.
Ellipses in the upper right corners of the frames indicate the spatial resolution.
Bottom left: broadband optical image (nearly Mould R) from \citet{koopmann2001}.
A smooth elliptical model of NGC 4649 has been subtracted from the image,
and contours are spaced by factors of two.
Bottom right: continuum-subtracted H$\alpha$ image from
\citet{koopmann2001}.
\label{tilefig}
}
\end{figure}

\begin{figure}
\epsscale{0.7}
\plotone{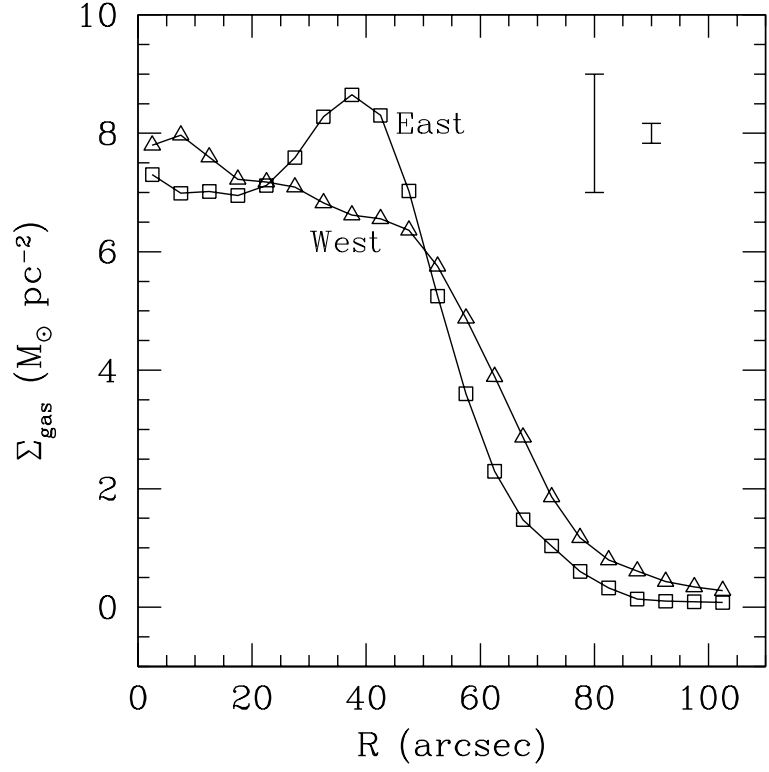}
\caption{Atomic gas column densities, averaged in semi-elliptical annuli.  
The east (receding) half of the galaxy is shown in squares, and the west
(approaching) half is in triangles.
The column densities are deprojected to face-on 
and are multiplied by a factor of 1.36 to account
for the presence of helium.
The larger error bar in the upper right corner indicates a typical 
magnitude of the  {\it dispersion about the mean} in each annular
semi-ellipse; the smaller error bar indicates
a typical magnitude of the {\it uncertainty in the estimated mean} for each
annulus.
These error bars are $\pm 1\sigma$, i.e. they are 2$\sigma$ in length.
\label{radialhi}
}
\end{figure}

\begin{figure}
\epsscale{0.7}
\plotone{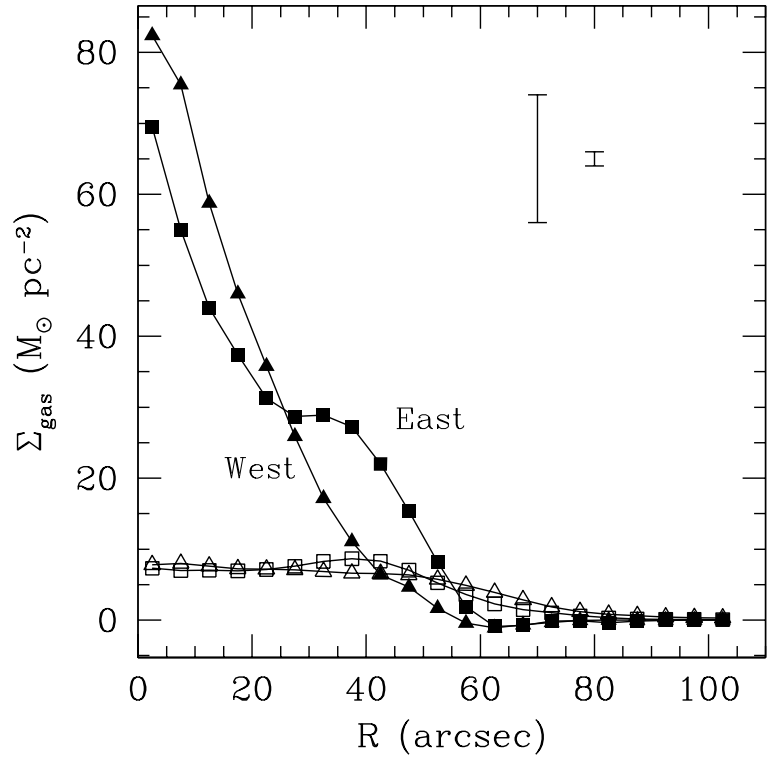}
\caption{Atomic and molecular column densities, averaged in
semi-elliptical annuli.  
The same HI column
density data that are shown in Figure \ref{radialhi} are reproduced
here in open symbols, and the molecular column densities are in filled
symbols.  In both cases, the east half of the galaxy is shown with squares
and the west half with triangles.
Typical 2$\sigma$ error bars for the molecular column densities are
indicated; see the caption to Figure \ref{radialhi} for explanation.
\label{radialcohi}
}
\end{figure}

The major features of the molecular distribution in NGC 4647 are a large
concentration near the center of the galaxy, a plateau on the east and a
steep radial dropoff to the southwest (Figures \ref{tilefig} and
\ref{radialcohi}).

The central molecular clump is poorly resolved by these 15\asec\ data, but
thirty percent (4.1\e{8} \solmass, including helium) of the molecular mass of the galaxy is
found within 15\asec\ of the central peak. At this radius the average
circular velocity of the two sides of the galaxy is 77 \kms\ (Section \ref{kinematics}), so the implied dynamical
mass is 1.7\e{9} \solmass\ or only four times the mass of the molecular
concentration.  
The central molecular clump thus makes a non-negligible contribution to the
dynamics in the interior of the galaxy.
The molecular peak is also 3\asec\ (250 pc) west of the $R$-band
nucleus, and this offset may have important consequences for the
gas kinematics in the nuclear regions.

The eastern half of the galaxy shows an extended molecular disk that is
coincident with the ring or arm segment seen in HI.  While the atomic gas
shows a 20\% enhancement in column density from one side to the other at
this 40\asec\ radius, the molecular gas shows a much stronger enhancement.
Molecular surface densities (corrected to face-on and including helium) are
30 M$_\odot$ pc$^{-2}$ in the arm/ring and are a factor of two to three
smaller on the west side  (Figure \ref{radialcohi}).
Thus the CO emission is strongly asymmetric and extends twice as far from
the galaxy center on the eastern side as on the western side.  

Because of the asymmetry in the CO distribution, the western side of the
galaxy shows a gradual decrease in atomic gas surface density but a very
sharp cutoff in molecular surface density.  The molecular distribution of
the galaxy then appears to have an edge at a position angle of 
roughly $-28\deg$.
This cutoff also makes itself visible in the spectra (Figure
\ref{spectra}), where it is plain that the CO does not extend far enough on
the low velocity side of the galaxy to sample the flat part of the rotation
curve.  The HI spectrum shows both ``horns" but the CO spectrum shows only 
one horn.

Star formation activity in NGC~4647 roughly follows the CO surface
brightness.  \halpha\ intensities are highest in the
nucleus of the galaxy, on the central CO clump, and are
lowest on the western side where there is little
molecular gas (Figure \ref{tilefig}).  
A few small \hii\ regions on the western side suggest that
there is probably a modest amount of molecular gas there but when averaged
over the 15\asec\ beam size it is below our sensitivity limits.
We also estimate the star formation efficiency (SFE) in the molecular gas by
convolving the continuum-subtracted \halpha\ image to the same resolution
as the CO image and dividing the two.  By this measure 
the SFE is not significantly different on the two
sides of the galaxy (at least over the regions where molecular gas is
detected).
The SFE in the galaxy nucleus is also comparable
to that in the disk, but the southeast
quarter of the galaxy has a SFE a factor of two to
three higher than the northeast quarter.  
Local molecular peaks in the arm of the galaxy tend to occur in between 
\hii\ regions, when measured on the scale of our CO resolution (1.2 kpc).
It would undoubtedly be useful to investigate possible effects of extinction on the
\halpha\ surface brightness through FIR or radio continuum images.

\subsection{Gas Kinematics}\label{kinematics}

\subsubsection{Asymmetry}

\begin{figure}
\includegraphics[scale=0.8,bb=40 235 588 584,clip]{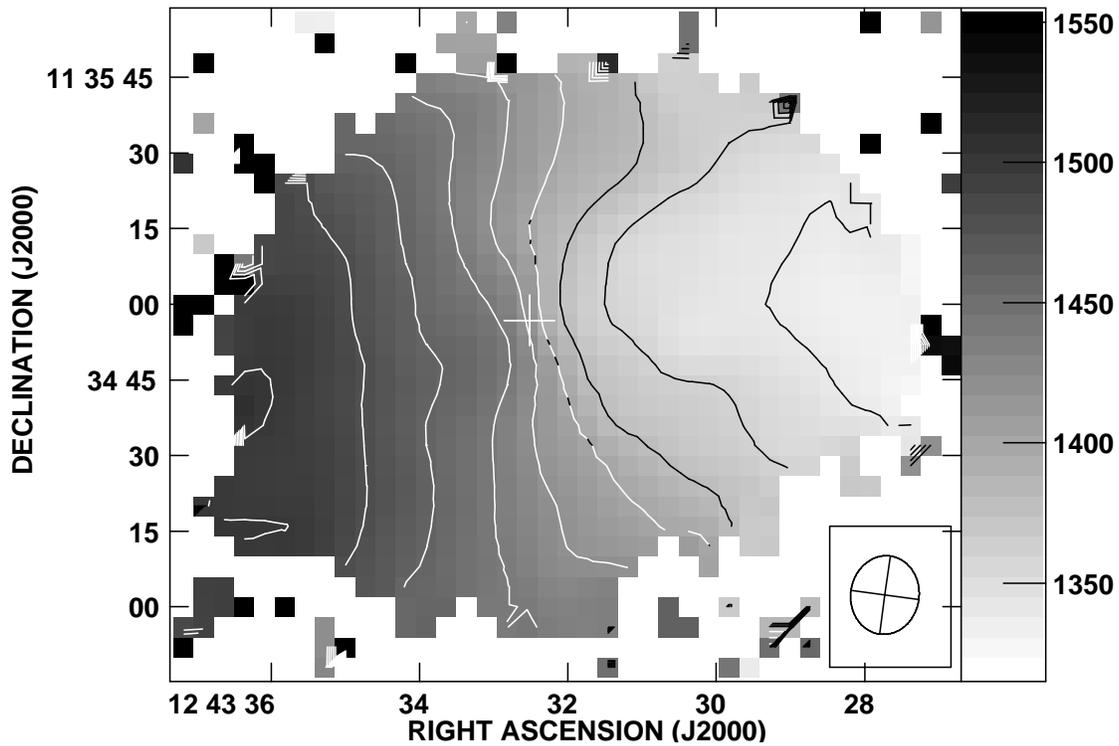} 
\caption{HI velocity field as indicated by the fitted center velocities at
each position.
The wedge on the right hand side of the figure indicates the
greyscale values,
and contours run from 1340 \kms\ to 1520 \kms\ in steps of 20 \kms.
The cross in the center of the figure indicates the adopted center of the
galaxy (the position of the nucleus in the R-band image).  The ellipse
in the bottom right corner shows the resolution of these data.
\label{HImom1}
}
\end{figure}

\begin{figure}
\includegraphics[angle=0,scale=0.8,bb=26 113 582 753,clip]{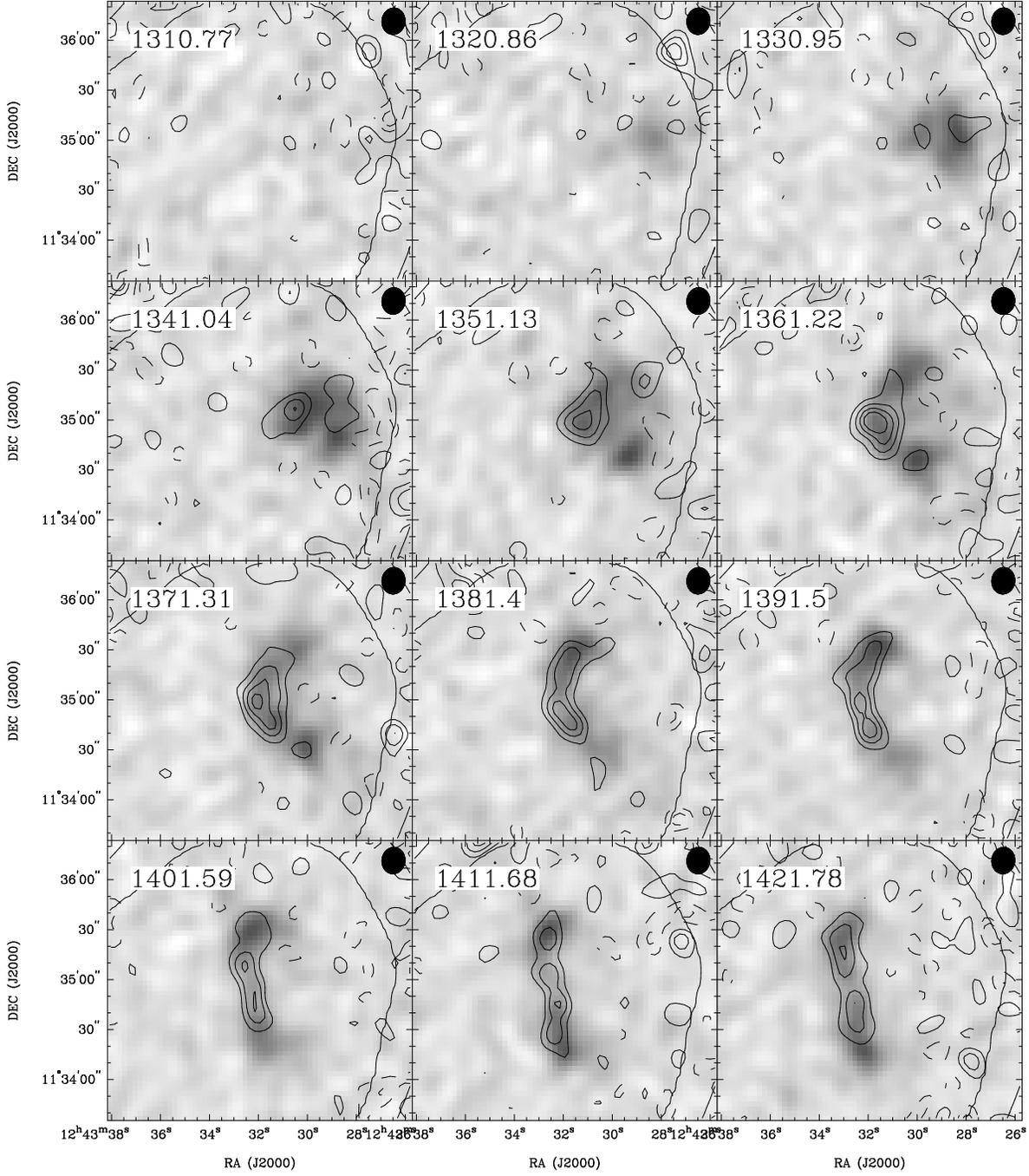}
\figurenum{7}
\caption{Individual channel maps.  HI emission is in greyscale.
Contours show CO emission at $-$4, $-$2, 2, 4, 6, and 8 times 
the rms noise level (100 \mjb).  The dark ellipse in every frame indicates
the angular resolution of both the HI and CO, and the velocity of each
channel (\kms) is indicated in the top left corner. 
A single large section of a roughly hexagonal contour indicates the points
where the gain of the CO image drops to 0.99, and small sections of the gain = 0.5
contour are visible in the extreme corners.
\label{channels}
}
\end{figure}

The atomic gas in NGC 4647 shows a striking asymmetry in its
kinematics (Figure \ref{HImom1}).
In the western (approaching) half, the side more distant
from NGC 4649, the rotation  curve rises quickly and flattens, giving a
typical spider-diagram pattern with isovelocity contours that bend toward
the major axis.  However, on the eastern side of the galaxy the velocity
contours are
much straighter and more suggestive of a linearly rising rotation curve.  This
pattern is also evident in the individual channel maps (Figure
\ref{channels}), which show the bent butterfly-wing structures on the west
side of the galaxy but linear features in the east.
Similarly, the major axis position-velocity diagram (Figure \ref{pv1}) shows 
that 
the west side rotation curve (RA offsets $< 0$) flattens whereas
the east side kinks at a radius of 10\asec\ and continues to rise.

\subsubsection{Kinematic parameters}\label{kinparm}

Analysis of the HI velocity field allows us to estimate the important
kinematic parameters of NGC 4647 and to quantify the kinematic asymmetries
discussed above.
We carried out a set of tilted ring fits with the {\it rotcur} task in the
GIPSY software package,
according to the method described by \citet{begeman89} and summarized
by \citet{swaters99}.

We first determine the position of the optical nucleus in NGC 4647 and
compare that to the kinematic center of the galaxy.
Ecliptic coordinates were transferred from a POSSII Digitized Sky Survey
plate to the nearly-Mould-R image of
\citet{koopmann2001} using seven stars distributed uniformly around NGC
4647 and NGC 4649.
The position of the galaxy nucleus in this image is 12\thr\ 43\tmin\ 32.58\tsec,
+11\arcdeg\ 34\arcmin\ 56.8\asec\ (epoch J2000) with an
uncertainty of half a pixel (0.4\asec).
A kinematic center is determined by first fitting all annular rings in the
galaxy with all of their parameters (center, position angle, inclination,
velocity) left free, and the average of all of the fitted centers gives a
kinematic center position only 2.5\asec\ away from the optical
position.  Additionally, a very simple fit of an inclined disk with a
rotation curve $V(r) \propto 1 - e^{r/r_0}$ (as encoded in the AIPS task
GAL) gives a kinematic center 4\asec\ away from the optical
nucleus.
As the resolution of the radio data
is 15\asec, we consider the kinematic centers to be consistent with
the position of the optical nucleus and we adopt the latter for the
dynamical center of the galaxy.  

Subsequent tilted ring fits with fixed centers show
significant trends in the systemic velocity from 
1407 \kms\ at radii less than 20\asec\ 
to 1425 \kms\ at 60--80\asec.
We attribute this trend to the asymmetry in the velocity field -- more
specifically, to the fact that the approaching side rotation curve flattens
whereas the receding side rotation curve keeps rising.
A value of 1410 \kms\ is adopted for the systemic velocity since it minimizes the differences between
the two sides at radii $< 25''$.

The kinematic position angle of the galaxy is
significantly better constrained than the systemic velocity.  Tilted ring
fits for position angle (with kinematic center fixed as described above) 
give position angles 98\deg -- 99\deg, and no trend with radius.
We also fit the model exponential velocity field to receding and approaching
sides independently and find that both sides of the galaxy give angles
consistent with the tilted ring fits.
We take the position angle of the kinematic major axis to be 98.5\deg \error\
0.5\deg, where the uncertainty indicates the dispersion between various
fits.  
Note, however, that this position angle is significantly different from
that adopted by \citet{rubin99} for longslit \halpha\ kinematics, which was 124\deg.  It is now clear that
the reason NGC 4647 appears to show minor axis rotation in the work of 
\citet{rubin99} is that their adopted kinematic minor axis is 30\deg\
offset from the true orientation-- we observe no minor axis rotation when the
proper kinematic position angles are used.

The kinematic major axis obtained from the HI velocity field is in good
agreement with the near-IR morphological major axis.
We obtained the Two Micron All-Sky Survey (2MASS)
K-band image of the NGC 4649-4647 system, masked out NGC 4647, and used the
Multi-Gaussian Expansion routines of \citet{cappellari02} to model NGC 4649
and remove it.  In subsequent MGE fits of NGC 4647 itself and isophote fits
using the ELLIPSE routine in STSDAS, we find the isophotal major axis 
position angles between 97.6\deg\ and 100.7\deg.

The receding half of the galaxy, with its almost linear isovelocity
contours,
gives very poor constraints on the inclination of the disk.  Therefore the
best information on the inclination comes from fits to the approaching half
alone, with fixed center and position angle.  Tilted ring fits are
consistent with the fit of the simple exponential velocity field and both
give inclinations 39.3\deg\ \error\ 1.8\deg.  The value is also consistent
with the inclination adopted by \citet{rubin99} as determined from optical
isophotes.
Figure \ref{rotcurve} shows the final rotation curve derived from the
tilted ring fits, where the receding and approaching sides were fit
independently (weighted by the square of the cosine of the angle to the
kinematic major axis).

The magnitude and spatial distribution of the kinematic asymmetry in 
NGC~4647 can be estimated by
taking the rotation curve derived from the approaching (flatter) side alone,
constructing a model velocity field, and subtracting from the
observed velocity field.  The residual velocities are on the order of 20 \kms\
at $r \sim 60''$ on the east side of the galaxy.  
If these residual velocities arise in tangential motions in the plane of the disk,
then correcting for inclination gives a 30 \kms\ velocity difference from one
side of the galaxy to the other (compared to a circular speed of 115 \kms).

\subsubsection{Atomic vs. Molecular and Ionized Gas Kinematics}

\begin{figure}
\includegraphics[scale=0.8,angle=-90]{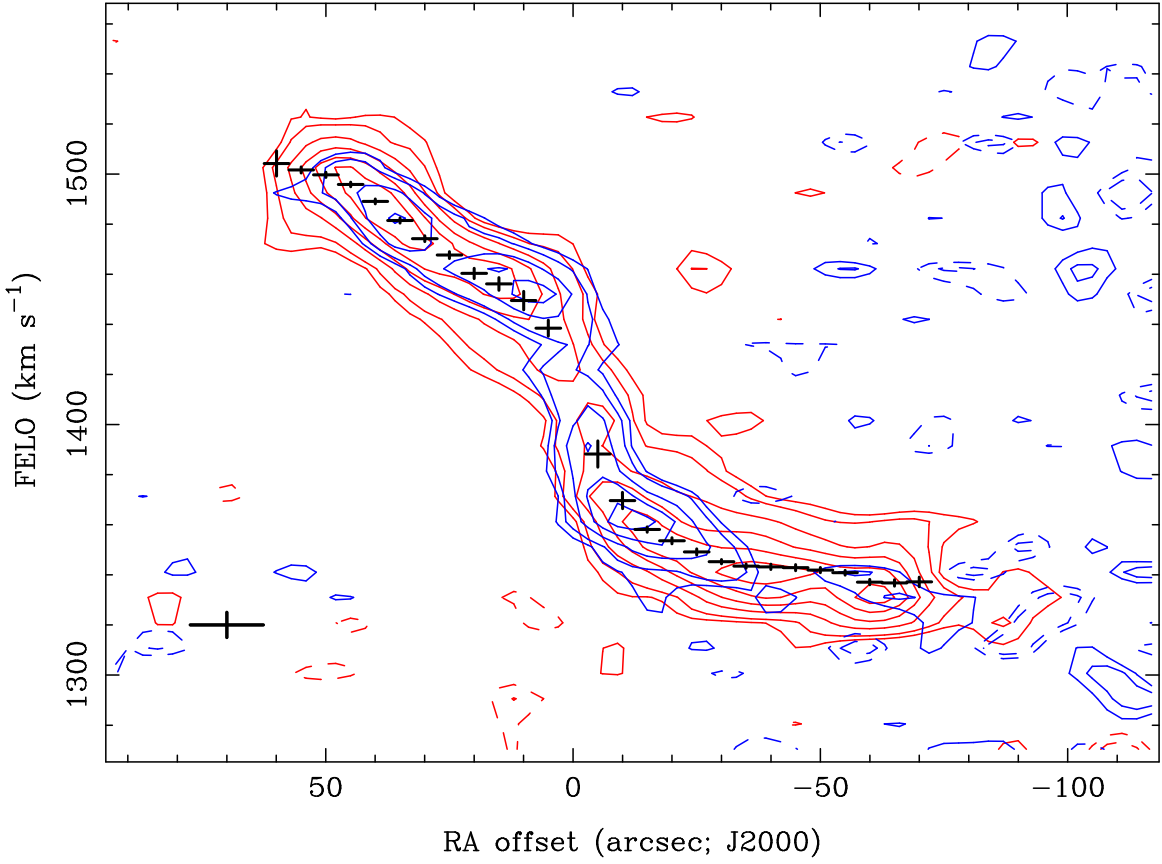}
\figurenum{8}
\caption{Major axis position-velocity diagram.  
HI contours are in red, at levels of $-3$, $-2$, 2, 3, 5, 7, 9, and 11
times the rms noise level (0.35 \mjb).   CO contours are in blue,
at levels of $-3$, $-2$, 2,3,5,7, and 9 times the rms noise level (100
\mjb).
The cross in the lower left corner indicates the velocity resolution
(channel width) and the spatial resolution (FWHM of the CO beam; the
resolution of the HI data is very similar).
Other crosses indicate the rotation curve derived from fitted tilted rings
to the two halves of the galaxy, as indicated in the text and 
as shown again in Figure \ref{rotcurve}.
The noise level of the CO data increases towards the right-hand
side of the figure because the sensitivity drops off toward the
edges of the mosaiced field of view.
\label{pv1}
}
\end{figure}

\begin{figure}
\epsscale{0.8}
\figurenum{9}
\plotone{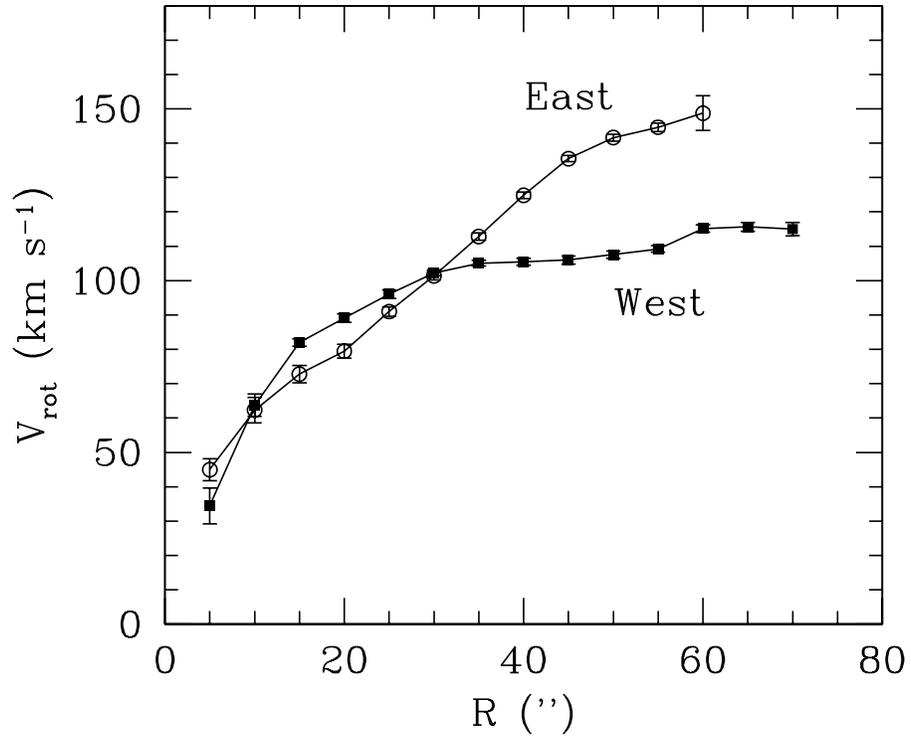}
\caption{HI rotation curve for the two sides fitted independently.
Open circles show the east side of the galaxy and filled circles show the
west side.
\label{rotcurve}
}
\end{figure}

\begin{figure}
\includegraphics[scale=0.6,angle=-90]{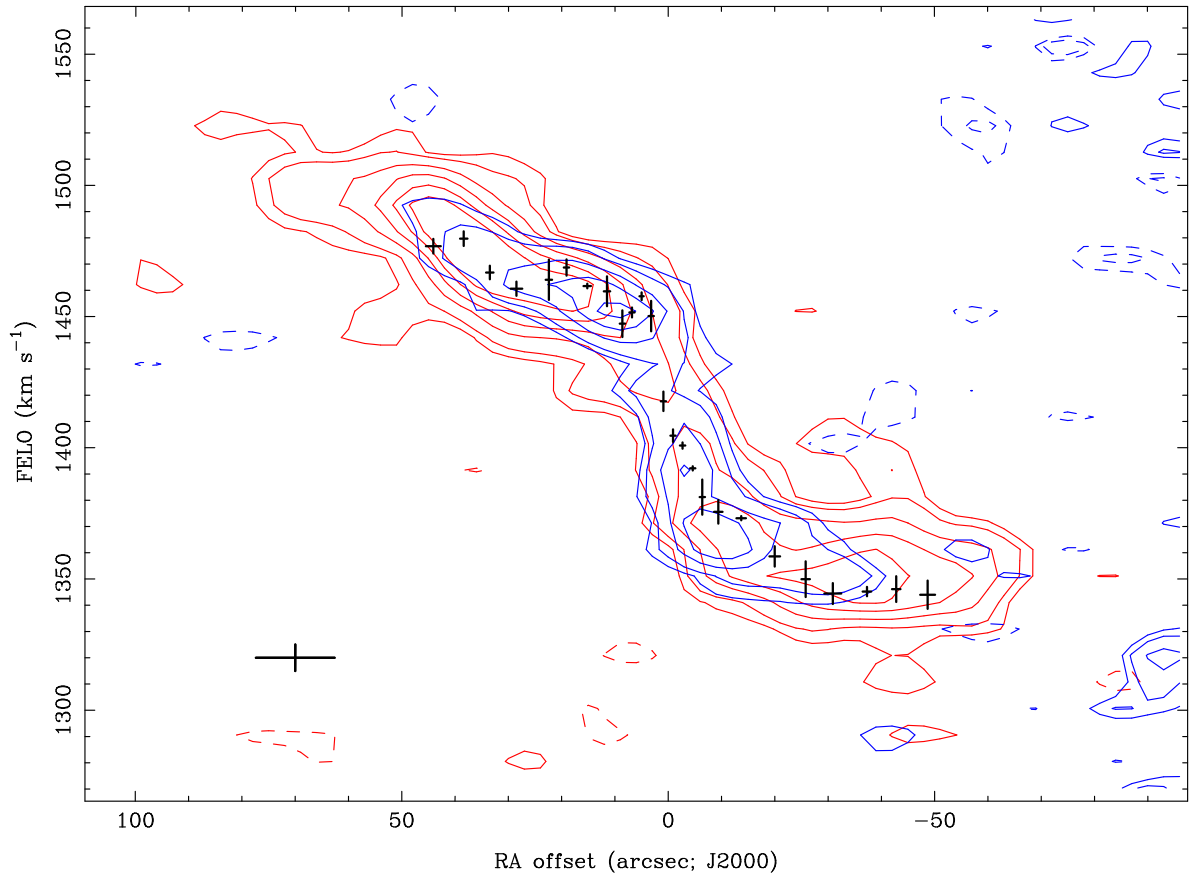}
\figurenum{10}
\caption{Position-velocity diagram at an angle of 124\deg.  This position
angle was the one adopted by \citet{rubin99} as the major axis.  
Contour levels and the resolution are the same as in Figure \ref{pv1}. 
The crosses here indicate the H$\alpha$ velocities measured by \citet{rubin99}.
\label{pv3}
}
\end{figure}

Molecular and ionized gas kinematics in NGC 4647 agree well with the atomic
gas.  For example, aside from the fact that the CO is more centrally
concentrated than HI and does not extend as far as HI on the western side, 
the channel maps in Figure \ref{channels} are nearly identical in CO and
HI.  The position-velocity diagrams in Figures \ref{pv1} and \ref{pv3} also
show good agreement between HI, CO, and \halpha\ kinematics.

There is one notable difference between molecular and atomic gas
kinematics: Figure \ref{pv1} shows that at radii between 10\asec\ and
30\asec\ on both sides of the nucleus the HI line profiles are asymmetric.
The line profiles have tails extending to smaller rotation speeds, 
which is symptomatic
of gas lagging the rotation curve by as much as 40 \kms\ (60 \kms, after
correcting for the inclination of the disk).
This HI ``beard" has also been observed in other spirals
\citep{fraternali01,barbieri05} and is interpreted as gas above
and below the plane of the disk.  We note that the molecular gas does not
show these asymmetric line profiles, so the gas that is inferred to be up
out of the plane is primarily atomic.

\subsubsection{Dispersions, Disk Instabilities, and Star Formation}
\label{veldisp}

Gaussian fits to individual line profiles are used to
estimate the velocity dispersions  of the HI and CO.
The contribution to the linewidth from beam smearing is estimated at each
point as the local dispersion in the velocity field (weighted by a Gaussian
of the same size as the spatial resolution).  Broadening by beam smearing
and finite channel width are then subtracted from the line width in
quadrature.  Aside from the nuclear regions where beam smearing dominates,
the corrected HI dispersions in the disk are 12\error\ 3 \kms; 
the latter quantity
indicates the width of the distribution rather than an uncertainty in
measurement.  The molecular gas has somewhat narrower lines, consistent with
dispersions 7 \error 3 \kms.
There is no measurable difference in the velocity dispersions on the two
sides of the galaxy.

The star formation rate in the disk may depend not only on the gas surface
density, but also on the degree to which the gas disk is gravitationally
unstable to the formation of molecular clouds \citep{krumholz05}.  The disk
instability is normally parametrized by Toomre's $Q$ parameter
\citep{Toomre64,BT},
$$ Q \propto \frac{\sigma \kappa}{G\Sigma_{gas}},$$
which is based on a comparison of the local sound speed $\sigma$ and the
epicyclic frequency $\kappa$  to the gas surface density $\Sigma_{gas}$.
Large values of the $Q$ parameter indicate stability against
the formation of self-gravitating entities.  

Given the differences in the
properties of the two sides of NGC 4647's disk, the galaxy makes an interesting test
case for the study of star formation.  The gas velocity dispersion is equal
on the two sides of the galaxy; however, the epicyclic frequency may not
be.  Under the assumptions that the HI and CO velocity fields
still trace the local circular velocity and that the inclination of the
disk is the same on both sides of the galaxy, we may use the results in
Figure \ref{rotcurve} to estimate the epicyclic frequency.
The side with the rising rotation curve has $\kappa =
2.1\times 10^{-15}$~\persec\ at a radius of 4 kpc, whereas the flatter side
of the rotation curve has $\kappa = 1.2\times 10^{-15}$~\persec\ at the same
radius.  
These initial estimates hint that the higher gas surface densities on the
east side of the galaxy may be offset by a higher epicyclic frequency on
that side, producing roughly equal values of Toomre's $Q$ parameter
on both sides.

\subsection{\SigHI, \SigHtoo, and Pressure}\label{pressure-sec}

The asymmetry in the CO intensity (Section \ref{comorph}) could conceivably
reflect either the underlying
distribution of molecular gas or an asymmetry in the CO luminosity per
unit \htoo\ mass.  
For example, a factor of two to three variation in the CO-to-H$_2$ 
conversion factor from one side of the galaxy to the other could reproduce
the observations while keeping the total gas surface density equal on both
sides.  This conversion factor is expected to depend on several properties of
the molecular clouds, including the gas density, temperature, and the
interstellar UV field to which the clouds are exposed.  Here we simply
comment that the UV field is probably stronger in the eastern half of the
galaxy than in the western half, since the \halpha\ luminosity is higher in
the east (Figure \ref{tilefig}).  A stronger UV field could either lead to
higher temperatures and increased CO luminosity \citep{Weiss2001} or to
preferential destruction of CO and lower CO luminosity \citep{Israel1997,
Lequeux1994}, and distinguishing between these alternatives is beyond the
scope of the present paper. 

We also note that if the conversion factor varies but the total gas surface
density is constant, then the star formation efficiency must be higher on
the eastern side of the galaxy in order to produce the higher \halpha\
intensities there (Section \ref{comorph}).  
A higher star formation efficiency on the eastern side might be in conflict
with the Toomre-style stability analysis of section \ref{veldisp}, which
showed that the epicyclic frequency is expected to be higher in the east.
Evidently the assumption of a variable conversion factor raises at least as
many questions as it answers.
Thus we proceed
under the assumption that the conversion factor is constant and the galaxy really does have factors of two to three
higher total gas surface density on the eastern side than on the western
side.

\citet{br04} propose that the H$_2$/\ion{H}{1} column
density ratio in a disk galaxy is determined by the hydrostatic
pressure.  This kind of a model could have important applications in
simulations of galaxy evolution, since it provides a way to calculate the
molecular surface density for star formation purposes.
The asymmetric gas distribution in NGC 4647 provides an
unusual opportunity to test this model.  From a different perspective, if
the model is correct then the observed column densities in NGC 4647 offer a
way to calculate the interstellar pressure on the two sides of the galaxy
and to find out which (if either) of the two sides is consistent with the
properties of other spirals.

We explore these questions by constructing a map of the hydrostatic
pressure in the galaxy using the method of \citet{br06}.  Their
formulation models disks of galaxies as two gravitationally-coupled
fluids with the stellar scale height and midplane volume density much
larger than those of the gas at any given position in the disk.  For
an infinite disk, the implied midplane pressure in the gas component
is
\begin{equation}
\frac{P_{ext}}{k} = 272 \mbox{ cm$^{-3}$ K}
\left(\frac{\Sigma_{g}}{M_{\odot}\mbox{ pc}^{-2}}\right)
\left(\frac{\Sigma_{\star}}{M_{\odot}\mbox{ pc}^{-2}}\right)^{0.5}
\left(\frac{v_g}{\mbox{km s}^{-1}}\right)
 \left(\frac{h_\star}{\mbox{pc}}\right)^{-0.5}.
\label{pext}
\end{equation}
where $\Sigma_{g}$ and $\Sigma_{\star}$ are the inclination corrected
surface densities of neutral gas and stars respectively.  The gas
velocity dispersion is given by $v_g$ and $h_\star$ is the stellar
scale height.  We convert the integrated intensity of the \ion{H}{1}
and CO maps into mass surface densities using the standard conversion
factor of $1.82\times 10^{18}~\mbox{cm}^{-2}/(\mbox{K km s}^{-1})$ for
the 21-cm emission and a CO-to-H$_2$ conversion factor of $2\times
10^{20}~\mbox{cm}^{-2}/(\mbox{K km s}^{-1})$.  The column densities
are then converted to mass surface density assuming a mean particle mass of
1.36 $m_{\mathrm{H}}$ per hydrogen nucleus and an inclination of
$39.3^{\circ}$ (\S\ref{kinparm}).  We convert the masked $K$-band
image of the galaxy to a stellar surface density assuming a
mass-to-light ratio of $M_K/L_K = 0.5~M_{\odot}/L_{\odot}$
\citep{bell-m2l}.  After convolving the H$_2$ and stellar surface
density maps to match the resolution of the \ion{H}{1} map, we aligned
the \ion{H}{1} and stellar surface density maps to the H$_2$ map.
Using the resulting maps of $\Sigma_\star$ and $\Sigma_g =
\Sigma_{\mathrm{HI}}+\Sigma_{\mathrm{H2}}$, the external pressure can
be calculated directly from Equation \ref{pext}, provided the gas
velocity dispersion and stellar scale height are known.  Following
\citet{br06}, we assume that a gas velocity dispersion of 8 km s$^{-1}$
appropriate for most galaxies and determine a stellar scale-height
(420 pc) from a fit to the stellar scale-length (3.2 kpc) using the
relationship of \citet{kregel-stellar}.

\begin{figure}
\epsscale{0.8}
\figurenum{11}
\plotone{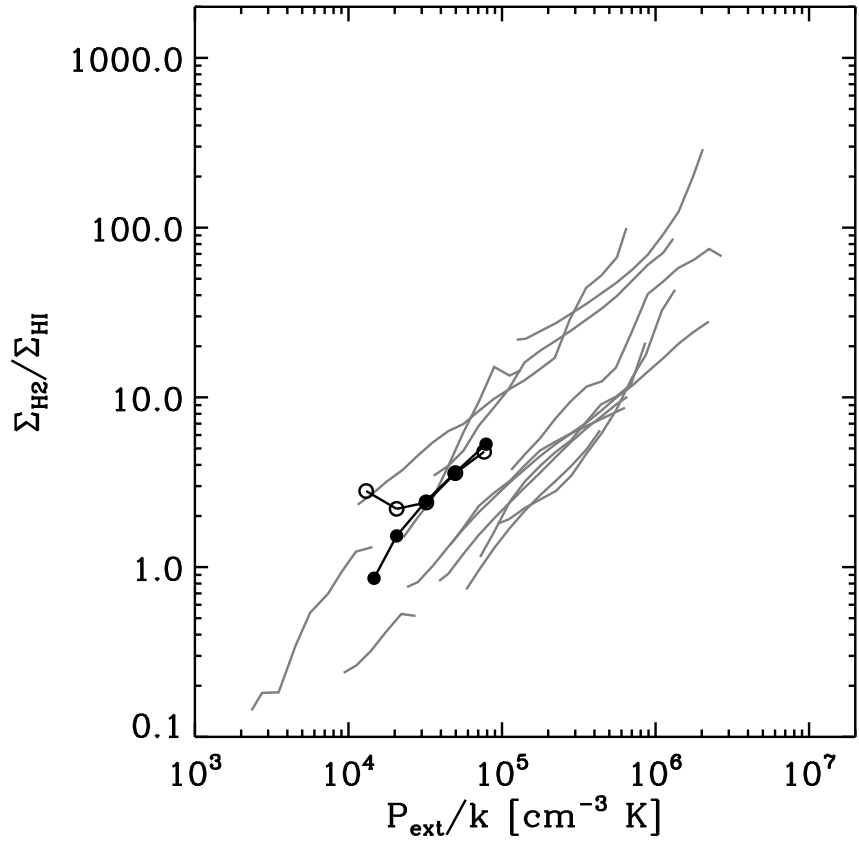}
\caption{H$_2$/\ion{H}{1} column density ratio vs.\ pressure.
The approaching (west) side of NGC 4647 is shown with filled circles and the
receding (east)
side with open circles.  The same relationships for 13 galaxies in
\citet{br06} are shown as gray lines.   For high pressures (near the
center of the galaxy), the molecular gas fraction agrees well in the
two halves.  The relationship appears to be the same as for other
galaxies.  However, at low apparent pressures, the column density
ratios diverge in the two halves. \label{pressure}
}
\end{figure}

In Figure \ref{pressure}, we plot the molecular gas fraction 
as a function of the midplane hydrostatic pressure $P_{ext}$
in the two halves of the galaxy (separated along the kinematic minor axis).
For the western half of the galaxy, the surface density ratio scales
with the hydrostatic pressure as it does in other galaxies
($\Sigma_{\mathrm{H2}}/\Sigma_{\mathrm{HI}}= (P_{ext}/P_0)^{1.1\pm
0.1}$).  The scale constant $P_0$ 
($=1.5\times 10^{4}~\mbox{K cm}^{-3})$ is consistent with
the other galaxies in the Virgo Cluster that were included in the
\citet{br06} sample.  In the eastern half of the galaxy, the surface
density ratio for low apparent pressures (large radii) is a factor of 3.3 higher
than in the western half.  In other words, the eastern half is
proportionally more molecule-rich than the western half for a given
apparent pressure in the ISM.  Provided that the pressure in the ISM does
indeed regulate the molecular gas fraction, this result implies that the
eastern half of the galaxy must be subject to larger pressures in the
ISM than are inferred from the mass components of NGC 4647 alone.  The
pressure deficit is $P/k=2.4\times 10^{4}~\mbox{K cm}^{-3}$, and this
deficit is twice as large as the midplane hydrostatic pressure in the
outer disk of the western side of the galaxy.

\section{Discussion}

In a $\Lambda$CDM universe galaxy-galaxy mergers and encounters are
recognized as major processes that have determined the spectrum of galaxy
properties.  From an observational perspective, galaxy asymmetries can serve
as indicators of the merger/interaction rate.  One may measure
the incidence of asymmetries and then make the leap to the
merger/interaction rate via several important assumptions such as 
the timescale over which asymmetries will remain detectable.
That timescale depends crucially on an understanding of the processes
responsible for the asymmetries and how they affect the gas, stars, and star
formation activity.

In this context our observations of the mildly disturbed galaxy NGC 4647
raise several questions.
NGC 4647 shows two kinds of lopsidedness, in its kinematics and in its gas
distribution.  More specifically, the gas distribution is asymmetric
in two ways:  the total gas surface density is significantly higher
on the east side than on the west, and the gas disk is also more highly
molecular on the east side than on the west.  How common are these types of
asymmetries, and what causes them?  We address the problems in two stages.
In section \ref{lopsided}, a comparison to other
mildly lopsided galaxies suggests processes that may create the kinematic asymmetry
and enhanced gas surface densities on the east side of the galaxy (or the
depressed surface densities on the west side).  
Section \ref{press-asym} discusses
local pressure enhancements that may drive the ISM into a
more highly molecular state on the east side of the galaxy.  In section
\ref{leftoverQ}  we 
list some remaining questions and propose tests that would give greater
insight into NGC 4647 in particular and spirals in general.

\subsection{Lopsided Galaxies}\label{lopsided}

It has been known for some time that many spirals are asymmetric or
lopsided.
For example, \citet{rz95} and \citet{zr97} find that some 30\% of field
spirals have lopsided stellar distributions. \citet{baldwin80} show
asymmetries in the HI distributions of spirals, and \citet{rs94} find that
50\% of field spirals show asymmetric global HI profiles that are probably
attributable to lopsided gas distributions.
\citet{rubin99} studied \halpha\ rotation curves for Virgo spirals,
and they find that half show significant kinematic disturbances.  In many
of these, the disturbance takes the form of an asymmetry in which the
velocity field is different on the two sides of the galaxy.
From studies of HI position-velocity diagrams and velocity fields
\citet{swaters99} estimate that the incidence of kinematic lopsidedness is
15\%--30\% among non-interacting field spirals.

The relatively high incidence of asymmetries
(even among field spirals) is significant, since it implies that the
responsible mechanisms must either be frequent or long-lasting.  Typical
dynamical timescales in spiral disks are only on the order of $10^8$ yr.
In this context,
four general processes have been proposed as sources of kinematic and/or
morphological lopsidedness: ram pressure in a cluster environment,
gravitational interactions between galaxies, asymmetric accretion of gas
from the cosmic web or from satellites, and asymmetric or off-center
gravitational potentials.

NGC 4647 is undoubtedly experiencing some ram pressure in its passage
through the Virgo intracluster medium.  Section \ref{press-asym} makes
quantitative estimates of the ram pressure.  At a purely qualitative level, 
the sharp HI edge on the northeast side and the shallow dropoff in the HI
column density on the southwest (Section \ref{HIdist}) are suggestive of
gentle stripping.  
The kinematic asymmetry in NGC 4647 may also be evidence of the non-circular
velocities induced by significant ram pressure \citep{hidaka2002}.
Finally, the form of the CO asymmetry in NGC 4647 is very like that in
NGC~4419, which was attributed to ram pressure by \citet{kenney1990}.

However, it is not at all obvious that ram pressure by itself is a
convincing explanation for the ensemble of properties displayed in NGC
4647.
The east side of the galaxy appears strongly disturbed, with its enhanced
CO emission and rising rotation curve.  At 40\asec\ 
radius the total gas density is twice as high on the east side
as on the west side.
In contrast, the west side appears very gently (if at all) disturbed.  Its
rotation curve is well behaved out to at least 70\asec.
Unless the galaxy is proceeding exactly edge-on through the intracluster
medium, so that the west side is shielded by the east side,
we might expect that the effects of ram pressure would be more noticeable on
the low density side of the galaxy than on the high density side.  
In this respect the ram-pressure stripped galaxy NGC~4419 is actually quite
different from NGC~4647; both have asymmetric molecular distributions, but
in NGC~4419 the atomic gas is very severely stripped as well
\citep{kenney04,chung06}.

Asymmetric gravitational potentials can also produce kinematic features like
those observed in NGC 4647 and other spirals.
For example, \citet{schoenmakers97} add a perturbation of the form
$\cos(m\theta)$ to the potential of an axisymmetric galaxy and calculate the
effects on its velocity field.
Based on that work \citet{swaters99} have shown that a $m=1$ perturbation in the 
potential, where the axis of the perturbation is aligned with the
kinematic major axis of the galaxy, can produce a velocity field that
flattens on one side but continues to rise on the other.  A 5\%--10\%
perturbation in the potential can induce 10\%--20\% amplitude differences in
the rotation curves of the two sides of the galaxy.
This kind of a lopsided potential is the explanation favored by
\citet{swaters99} to explain the HI velocity fields of DDO~9 and NGC~4395,
which are similar in character to NGC 4647.
\citet{noordemeer01} have also shown that the lopsided kinematics of NGC~4395
(and, by analogy, NGC~4647) could be explained by setting the galaxy disk
off-center with respect to the dark matter halo.  \citet{battaglia06} also
propose that the halo of NGC~5055 is offset from the center of its disk.

A lopsided gravitational potential would affect the stellar distribution in
NGC~4647 in addition to the gas distribution.  There are subtle
hints of a $m=1$ asymmetry in the stellar distribution of the galaxy; for
example, in the broadband red image of Figure \ref{tilefig} the outermost
contours are compressed on the southern side of the galaxy and extended on
the northwest.  The galaxy nucleus is therefore not quite centered in the
middle of the outer contours.  NGC 4649 has been modeled and subtracted,
but more careful analysis should be done to quantify any degree of
asymmetry in the stellar distribution of NGC 4647.

NGC~4647, of course, is also an obvious candidate for a gravitational
interaction with NGC~4649.
\citet{bournaud05} have
modelled distant, high velocity 
encounters of similar-mass galaxies, and they find that for impact parameters
130--450 kpc and velocities 160--450 \kms, the retrograde 
in-plane encounters can make $m=1$ asymmetries in which the ratio of the
perturbed to the unperturbed gravitational force is greater than 0.1 for several Gyr.
These amplitudes are similar to those required by \citet{swaters99}.

A statistical analysis of the distribution of galaxy lopsidednesses leads
\citet{bournaud05} to propose that asymmetric accretion of gas from the
cosmic web must be a significant contributor to lopsidedness in general.
The mechanism simply requires that the incoming gas have a nonzero impact
parameter.
They find that if accretion rates are fairly high --- several
\solmass~\peryr, or enough to double the galaxy masses over a Hubble time
--- a galaxy may develop an $m = 1$ asymmetry with an amplitude of 10\% to
20\%.  The asymmetries in the gas and stellar distributions can persist
even for several Gyr after accretion stops.  As NGC~4647 is embedded in the
Virgo cluster's hot ICM it is unclear how this process
might apply to NGC~4647, what the phase of the accreted gas would be, or
whether the effect on galaxy kinematics would match the observations.

We suspect that the process that is responsible for the kinematic
asymmetry also enhances the gas densities on the eastern side of the
galaxy.  But the opposite perspective is also interesting: the pile-up in 
gas surface density may contribute to the kinematic asymmetry.
We estimate the dynamical mass of the galaxy taking a rotation speed of 115
\kms\ at a radius of 70\asec\ on the approaching (flatter, probably less
disturbed) side of the 
rotation curve; these values give 1.8\e{10} \solmass\ within 5.8 kpc.
Some 7.7\% of this dynamical mass is attributable to molecular gas plus
helium.  Since most of the molecular gas is found on the east side of the
galaxy, that gas distribution may well contribute to a few percent
perturbation in the gravitational potential.   The gas density asymmetry is also a
$m=1$ perturbation whose axis is roughly aligned with the kinematic major axis, so
it has the correct form to explain the kinematic asymmetry according to the
models of \citet{swaters99}.
Therefore, even if the original cause produced both the kinematic and the gas
density asymmetries together, the resulting density asymmetry may help to
maintain the kinematic asymmetry over timescales longer than a dynamical
time.

\subsection{Producing the Pressure Asymmetry}\label{press-asym}

If pressure regulates the chemical state of the ISM, as suggested by
\citet{br06}, then the asymmetry in molecular gas fraction is
produced by an asymmetry in ISM pressure in the disk.  Pressure, by
itself, cannot explain the enhanced gas column density on the east side
of the galaxy.
However, once the gas is piled up on one side, pressure could explain the
amount of the ISM that is molecular.  The analysis of section
\ref{pressure-sec} finds that the observed asymmetry in the molecular
gas fraction would imply an excess pressure of $P/k=2.4\times
10^{4}~\mbox{K cm}^{-3}$ beyond that required for the hydrostatic
support of the disk of NGC 4647.  
Thus, we consider again the effects of the intracluster medium, the proximity
of NGC 4649, and asymmetric gravitational potentials as possible sources of
extra pressure on the east side of NGC 4647.

At the $\sim 1$ Mpc separation between M87 and NGC 4647, the
intercluster medium (ICM) of Virgo has a density of $n\sim
10^{-4}~\mbox{ cm}^{-3}$ and a temperature of $\sim 2$ keV
\citep{kenney89}. The isotropic pressure in the ICM is significant:
$P_{ICM}/k\sim 2\times 10^{3}\mbox{ cm}^{-3}\mbox{K}$, on the order of 10\%
of the midplane hydrostatic pressure in the outer parts of NGC 4647's disk.
If NGC 4647
is moving with respect to the ICM, it suffers additional ram pressure  
\begin{equation}
P_{ram} = \frac{1}{2}\mu m_{\mathrm{H}} n v^2 = 1.3\times 10^{4}~
\mbox{cm}^{-3}\mbox{ K} \left(\frac{v}{800 \mbox{ km s}^{-1}}\right)^2.
\end{equation}
The equation has been normalized to the one-dimensional velocity
dispersion of the Virgo Cluster \citep{huc85}.  NGC 4647 has a
line-of-sight motion relative to M87 of only 100 km s$^{-1}$, but if its
tangential velocity carries it eastward on the sky at 1000 \kms\ or so
then the ram pressure
could be strong enough to effect the inferred pressure difference. 
It remains to be seen whether the extra pressure could be applied delicately
enough to avoid major disruptions in the kinematics of the downstream side of the galaxy.

Provided that NGC 4647 is physically close to NGC 4649, the hot gas in NGC
4649 can increase the pressure on the eastern side of NGC 4647 through two
mechanisms.
The first is ram pressure between NGC 4647 and the halo of NGC 4649.  
Indeed, the diffuse X-ray emission from NGC 4649 extends out to the
projected distance of NGC 4647 with a suggestion of a bow shock between the
two \citep{randall04}.
If the galaxies are approaching each other, a lower limit on the
relative velocities of the gas systems is 
$V_{4647}+V_{rot}\cos i-V_{4649}$=400 km s$^{-1}$.  
An enhancement in the plasma density of 0.002 particles cm$^{-3}$ on
the eastern side of NGC 4647 would produce the necessary pressure to
account for the observed asymmetry.  Using the density profile of the
X-ray halo of NGC 4649 \citep{bri97}, we estimate that this particle
density occurs at roughly 12 kpc from the center of NGC 4649.
The ISM of NGC 4647 may also be compressed by the isotropic
pressure of the hot halo gas; again, the thermal pressure
of NGC 4649 is $2.4\times 10^4\:\mbox{cm}^{-3}$ at a distance of $\sim
12$ kpc from the center of the galaxy \citep{bri97}.
The two galaxies would have to be no farther apart than their current
projected separation for these mechanisms to drive the molecular asymmetry.

The difficulty with appealing to a close approach of NGC 4647 and NGC 4649 to
explain the kinematic and pressure asymmetries is, again, it seems unlikely
that the disk of NGC 4647 could survive the tidal forces and remain unscathed.
We take the mass of NGC 4649 to be at least $10^{12}$ \solmass\ based on the
total $V$ magnitude and a $V$-band mass-to-light ratio of 9.5--16
\citep{debruyne01}.  Strictly speaking, the dynamical analysis of
\citet{debruyne01} measures the mass-to-light ratio within the effective
radius so that this dynamical mass of $10^{12}$ \solmass\ is probably an
underestimate.  For galaxy separations as small as 30
kpc, the tidal acceleration at the edge of NGC 4647's HI disk is two-thirds as
large as the spiral's own centripetal acceleration.
We therefore consider it extremely unlikely that NGC 4647 and NGC 4649 are
presently receding from a past encounter at a few tens of kpc.  It may be
possible that they are still approaching each other.

Alternatively, if something (NGC 4649, or an off-center dark matter halo)
generates a mild
$m=1$ mode in the potential of NGC 4647 as described in section \ref{lopsided},
the gas must speed up and slow down as it executes its nearly circular
orbit.  The ram pressure of gas
flowing down into the trough in the potential may be sufficient to account for the
expected asymmetry in pressure.  This possibility is especially
appealing since it naturally connects the asymmetries in the velocities, the 
gas surface density, and the molecular fraction.
The ram pressure of such a flow is
\begin{equation}
P_{ram} = \frac{1}{2}\mu m_{\mathrm{H}} n v^2 = 3\times 10^{4}~
\mbox{cm}^{-3}\mbox{ K} \left(\frac{n_{\mathrm{H2}}}{\mbox{ cm}^{-3}}\right)
\left(\frac{v_{flow}}{10 \mbox{ km s}^{-1}}\right)^2
\end{equation}
where $v_{flow}$ is the relative velocity at which gas approaches
the potential minimum.  
For flow velocities comparable to the observed velocity differences on the
two sides of the galaxy, the
combination of the ram pressure with the ICM and the pressure of the
gas accumulating in the $m=1$ potential could explain the observed gas
distributions.

Finally, since the implied hydrostatic pressure scales linearly with the gas
velocity dispersion, we note that the molecular asymmetry could be explained
without recourse to an additional source of pressure if the gas velocity
dispersion is higher on the east side of the galaxy than on the west.
However, to account for the molecular asymmetry entirely, the velocity dispersion
would have to be $>20$ \kms\ on the east side.
Such high dispersions are ruled out by our observations (section \ref{veldisp}). 

\subsection{Remaining Questions}\label{leftoverQ}

How long the asymmetries in the kinematics and the gas distribution of NGC
4647 will last depends critically on their causes.  If the disturbance is
temporary, such as a fly-by of NGC 4649, then the spiral can be
expected to settle back to a more symmetric state eventually.  Alternatively,
an offset dark matter halo might produce a long-lasting asymmetric state.
The discussions above do not provide a definitive answer to the source of the
asymmetries in NGC 4647, but a variety of observational and simulation-based
tests can help confirm or rule out various processes.

Part of the difficulty with understanding NGC 4647 is that the separation
between NGC 4647 and NGC 4649 is still unknown (as is the true relative
velocity of the two).  However, if the asymmetries of NGC 4647 are due to a
galaxy-galaxy interaction then we would expect the stellar
kinematics of NGC 4649 to reflect that interaction.  N-body simulations could
be used to predict the character and the magnitude of those interaction
signatures.

We have also discussed whether hydrodynamic processes such as ram pressure
stripping or off-center gas accretion could be responsible for the kinematic
and gas distribution features of NGC 4647, but we do not know the history of
NGC 4647's interaction with the ICM.  The importance of hydrodynamic
processes could be assessed by comparing the gas kinematics to the 
kinematics of the old stellar populations.  If the stellar kinematics are
symmetric, then the asymmetries in the gas can probably be attributed to
temporary hydrodynamic effects.
In addition, hydrodynamic simulations could be used to 
test whether the ensemble of properties displayed by
NGC~4647 implies the presence of both ram pressure and gravitational
disturbances (as \citet{vollmer03} has suggested for NGC 4654).

A comparison of the $m=1$ asymmetry in the gas surface density to the models
of \citet{swaters99} suggested that the lopsided gas distribution might 
contribute to the observed kinematic asymmetry in the gas.  That idea could be tested
with appropriate hydrodynamic simulations. 
In addition, simulations that can follow heating, cooling,
and formation and dissociation of molecules could
investigate whether the conversion from HI into \htoo\ follows
naturally from the differences in the gas velocity from one side to the
other.

H$\alpha$ images of NGC 4647 are noticeably asymmetric (Figure
\ref{tilefig}), and we have shown that the H$\alpha$ surface
brightness roughly traces the CO surface brightness.  If the \htooco\
conversion factor is roughly constant throughout the galaxy, the
star formation efficiency is constant as well.
Yet the epicyclic frequency, an important component of a local
gravitational stability analysis, may be significantly different on the two
sides of the galaxy.
More detailed analyses of the stability of this asymmetric
disk could give valuable insight into large-scale star formation processes.

The pressure model of \citet{br04} is consistent with our observations of NGC
4647, to the extent that there are several viable candidate sources for the
implied extra pressure on the east side of the galaxy.
Further progress in testing the model could be stimulated by a better
knowledge of how many spirals show relatively symmetric atomic gas but
strongly asymmetric molecular gas.

We note that NGC 4647
is actually classified as kinematically regular (non-disturbed) by
\citet{rubin99}.  Its kinematic asymmetries are indeed milder than those of
many Virgo spirals, as one can see by perusing the rotation curves
presented by those authors.  But the asymmetries of NGC 4647 are more obvious in 
two-dimensional velocity field data than in one-dimensional longslit data.
The properties of NGC 4647 therefore suggest that
the true incidence of disturbances among Virgo spirals is higher
than the 50\% measured by \citet{rubin99}.  

Furthermore, NGC 4647 would probably not have been identified as asymmetric on the
basis of its global HI profile alone (Figure \ref{spectra}).  The galaxy
therefore suggests that the incidence of asymmetric gas distributions may
be higher than the 50\% rate inferred by \citet{rs94}.  It is only the
combination of the HI and the CO data that revealed the true peculiarities
in its gas distribution.  
Likewise, it is the combination of these two gas phases that will give the
tightest constraints on the galaxy's history.
Additional matched resolution
HI and CO maps of Virgo spirals would be very helpful for understanding the
evolution of the ISM in mildly disturbed galaxies.

\section{Summary}

Images of HI and CO emission at 15\asec\ resolution
in the Virgo Cluster spiral NGC 4647 show that the
neutral gas is strongly lopsided. 
A spiral arm or a segment of a ring on the east side of the
galaxy shows up as enhanced
HI column densities and a plateau in the molecular gas distribution.  The
west side shows a very steep dropoff in molecular content but atomic gas
dribbles out to large radii. Thus, at radii larger than 30\asec\ (2.5 kpc)
the ISM is much more highly molecular on the east side of the galaxy than
on the west side.  As a result, total gas surface densities are twice as
high on the east side of the galaxy as on the west side.
The galaxy also shows a kinematic asymmetry; the rotation curve
flattens on the west side of the galaxy but continues to rise on the east
side.  The velocity difference between the two sides is nearly 30\% at the
edge of the HI distribution.

An analysis of the surface densities of atomic gas, molecular gas, and the
stellar disk suggests that the enhanced molecular fraction on the east side
can be explained if the interstellar pressure is higher there than in the
west.  The implied pressure difference is about a
factor of two (2.4\e{4} cm$^{-3}$~K) in the outer regions of the disk.
We conclude that the pressure model of \citet{br04} is probably consistent with the
gas distribution in NGC 4647 since there are several different plausible
mechanisms that could provide such a pressure difference.

We discuss possible causes of the kinematic and gas distribution
asymmetries in NGC 4647, focusing mainly on ram pressure and on asymmetries
in the gravitational potential.  The facts that (1) the gas distribution asymmetry is
strong in the molecular gas but weak in the atomic gas and (2) the galaxy
kinematics are less disturbed on the side where the total gas density is a
factor of two lower lead us to suggest that ram pressure by itself may not be
a satisfactory explanation for the gas pileup on the east side.
In addition, attributing
the pressure difference directly to NGC 4649
requires the two galaxies to be so close that tidal forces should destroy the
disk of NGC 4647.
Thus we favor attributing both the
kinematic and the morphological asymmetries to a lopsided gravitational
potential that may or may not be related to the presence of NGC 4649.
Gas settling into a potential well might explain the larger velocities on
the east side, and shocks could convert the gas to molecular form.

We also note several remaining questions that are raised by these results and
suggest directions for future work.
For example, observations of the stellar kinematics of NGC 4649 should help
test for a gravitational interaction between the two galaxies.  Stellar
kinematics of NGC 4647 could distinguish whether the forces disturbing its
gas are gravitational or hydrodynamical in origin.  Hydrodynamic simulations
could test whether the kinematic asymmetry drives the large-scale variation
in the molecular fraction of the gas.  A better understanding of
the cause(s) of the disturbance will help to predict how long the
disturbances will remain detectable.

The comparison of the low density atomic medium and the high density
molecular medium is vital to obtaining a complete picture of the ISM in
this mildly disturbed galaxy.  Furthermore, all indications are that a
large fraction of spiral galaxies (both in clusters like Virgo and in the
field) have kinematic disturbances like those in NGC 4647.  Additional
comparisons of atomic and molecular gas in mildly disturbed galaxies should
offer important insights into the evolution of the molecular phase as
galaxies settle into groups and clusters.

Finally, star formation activity (traced by \halpha\ emission) follows the
CO intensity very well on scales larger than about a kpc.
By this measure the local star formation efficiency is constant throughout the galaxy to
within factors of two to three.

\acknowledgments 

This paper made use of the Digitized Sky Surveys, 
produced at the Space Telescope Science
Institute under U.S. Government grant NAG W-2166. 
The images of
these surveys are based on photographic data obtained using the Oschin Schmidt 
Telescope on Palomar Mountain and the UK Schmidt Telescope. The plates were
processed into the present compressed digital form with the permission of these
institutions.
H$\alpha$ and R images of NGC 4647 were obtained through the
NASA/IPAC Extragalactic Database (NED), which is operated by the Jet Propulsion Laboratory, California Institute of Technology, under contract with the National Aeronautics and Space Administration. 
This work has been supported by National Science Foundation grants AST
0074709 and AST 0502605.
Thanks also to M.\ Cappellari for assistance with his MGE decomposition
routines.

\includegraphics[angle=0,scale=0.8,bb=26 113 582 753,clip]{f7b.eps}
\centerline{Fig. 7. --- Continued.}

\begin{thebibliography}{} 

\bibitem[Akeson(1998)]{akeson98}Akeson, R. 1998, BIMA memo series \#68

\bibitem[Baldwin, Lynden-Bell, \& Sancisi(1980)]{baldwin80} Baldwin, J.~E., Lynden-Bell, D., \& Sancisi, R.\ 1980,
\mnras,  193,  313 

\bibitem[Barbieri et al.(2005)]{barbieri05} Barbieri, C.~V., Fraternali,
F., Oosterloo, T., Bertin, G., Boomsma, R., \& Sancisi, R. 2005, \aap, 439,
947

\bibitem[Battaglia et al.(2006)]{battaglia06} Battaglia, G., Fraternali, F.,
Oosterloo, T., \& Sancisi, R. 2006, \aap, 447, 49

\bibitem[Begeman(1989)]{begeman89} Begeman, K.~G. 1989, \aap, 223, 47

\bibitem[Bell \& de Jong(2001)]{bell-m2l} Bell, E.~F., \& de
Jong, R.~S.\ 2001, \apj, 550, 212
                                                                                      
\bibitem[Binney \& Tremaine(1987)]{BT} Binney, J.~S., \& Tremaine, S. 1987

\bibitem[Blitz \& Rosolowsky(2004)]{br04} Blitz, L., \&
Rosolowsky, E.\ 2004, \apjl, 612, L29
                                                                                      
\bibitem[Blitz \& Rosolowsky(2006)]{br06} Blitz, L., \&
Rosolowsky, E.\ 2006, ApJ, accepted (astro-ph/0605035)
                                                                                      

\bibitem[Bournaud et al.(2005)]{bournaud05} Bournaud, F., Combes, F., Jog,
C.~J., \& Puerari, I. 2005, \aap, 438, 507

\bibitem[Brighenti \& Mathews(1997)]{bri97} Brighenti, F., \&
Mathews, W.~G.\ 1997, \apjl, 486, L83

\bibitem[Briggs, Schwab, \& Sramek(1999)]{SIRA2} Briggs, D.~S., Schwab,
F.~R., \& Sramek, R.~A. 1999, in Synthesis Imaging in Radio Astronomy II,
eds. G. B. Taylor, C. L. Carilli, and R. A. Perley (ASP Conf. Ser., Vol.
180), pp 127--150

\bibitem[Cappellari(2002)]{cappellari02} Cappellari, M. 2002, \mnras, 333,
400

\bibitem[Cayatte et al.(1990)]{Cayatte90} Cayatte, V., van Gorkom, J.~H.,
Balkowski, C., \& Kotanyi, C. 1990, \aj, 100, 604

\bibitem[Chung et al.(2006)]{chung06} Chung, A., et al. 2006, in
preparation (see also http://www.astro.yale.edu/viva)

\bibitem[De Bruyne et al.(2001)]{debruyne01} De Bruyne, V., Dejonghe, H., Pizzella, A., Bernardi, M., \& 
Zeilinger, W.~W.\ 2001, \apj,  546,  903 

\bibitem[de Vaucouleurs et al.(1991)]{RC3}
de Vaucouleurs, G., de Vaucouleurs, A., Corwin Jr., H.G., Buta, R.~J.,
Paturel, G., \&  Fouque, P. 1991,
Third Reference Catalogue of Bright Galaxies, Version 3.9
                                                                                

\bibitem[Elmegreen(1993)]{elmegreen93} Elmegreen, B.~G.\ 1993,  \apj,  411,  170 

\bibitem[Fraternali et al.(2001)]{fraternali01} Fraternali, F., Oosterloo,
T., Sancisi, R., \& van Moorsel, G. 2001, \apjl, 562, 47

\bibitem[Gavazzi et al.(1999)]{gavazzi99} Gavazzi, G., Boselli, A., Scodeggio, M., Pierini, D., \& Belsole, E.\ 
1999,  \mnras,  304,  595 
 
\bibitem[Helou, Hoffman \& Salpeter(1984)]{helou1984} Helou, G., Hoffman,
G.~L., \& Salpeter, E.~E. 1984, \apjs 55, 433

\bibitem[Hidaka \& Sofue(2002)]{hidaka2002} Hidaka, M., \& Sofue, Y. 2002,
\pasj, 54, 33

\bibitem[Huchra(1985)]{huc85} Huchra, J.~P.\ 1985, ESO 
Workshop on the Virgo Cluster, 181 

\bibitem[Israel(1997)]{Israel1997} Israel, F.~P. 1997, \aap, 328, 471

\bibitem[Jarrett et al.(2003)]{2003AJ....125..525J} Jarrett, T.~H.,
Chester, T., Cutri, R., Schneider, S.~E., \& Huchra, J.~P.\ 2003, \aj, 125,
525

\bibitem[Kauffmann et al.(1993)]{kauffmann93} 
Kauffmann, G., White, S.~D.~M., \& Guiderdoni, B.\ 1993, 
\mnras, 264, 201

\bibitem[Kenney et al.(2004)]{kenney04} Kenney, J.~D.~P., Crowl, H., van Gorkom, J., \& Vollmer, B.\ 2004,  
Spiral Galaxy - ICM Interactions in the Virgo Cluster,  IAU Symposium  
217,  370 

\bibitem[Kenney \& Young(1988)]{kenney88} Kenney, J.~D., \& Young, J.~S.
1988, \apjs, 66, 261

\bibitem[Kenney \& Young(1989)]{kenney89} Kenney, J.~D., \& Young, J.~S.
1989, \apj, 344, 171

\bibitem[Kenney et al.(1990)]{kenney1990} Kenney,
J.~D.~P., Young, J.~S., Hasegawa, T., \& Nakai, N. 1990, \apj, 353, 460

\bibitem[Koopmann \& Kenney(2004)]{koopmann2004} Koopmann, R., \& Kenney,
J.~D.~P. 2004, \apj 613, 866

\bibitem[Koopmann, Kenney, \& Young(2001)]{koopmann2001} Koopmann, R.,
Kenney, J., \& Young, J.~S. 2001, \apjs, 135, 125

\bibitem[{{Kregel} {et~al.}(2002){Kregel}, {van der Kruit}, \& {de
  Grijs}}]{kregel-stellar}
{Kregel}, M., {van der Kruit}, P.~C., \& {de Grijs}, R. 2002, \mnras, 334,
646

\bibitem[Krumholz \& McKee(2005)]{krumholz05} Krumholz, M.~R., \& McKee,
C.~F. 2005, \apj, 630, 250

\bibitem[Lay(1999)]{lay99}Lay, O. 1999, BIMA memo series \# 72

\bibitem[Lequeux et al(1994)]{Lequeux1994} Lequeux, J., Le Bourlot, J.,
Pineau des For\^ets, G., Roueff, E., Boulanger, F., \& Rubio, M. 1994,
\aap, 292, 371

\bibitem[Miller \& Owen(2003)]{miller03a} Miller, N.~A., \& Owen, F.~N.\ 2003,  
\aj,  125,  2427 
 
\bibitem[Miller, Owen, \& Hill(2003)]{miller03b} Miller, N.~A., Owen, F.~N., \& Hill, J.~M.\ 2003,  
\aj,  125,  2393 
 

\bibitem[Noordermeer et al.(2001)]{noordemeer01} Noordermeer, E., 
Sparke, L.~S., \& Levine, S.~E.\ 2001, \mnras, 328, 1064 

\bibitem[Randall et al.(2004)]{randall04} Randall, S.~W., Sarazin, C.~L., \&
Irwin, J.~A. 2004 \apj 600, 729.

\bibitem[Richter \& Sancisi(1994)]{rs94} Richter, O.-G., \& Sancisi, R.
1994 \aap\ 290, L9

\bibitem[Rix \& Zaritsky(1995)]{rz95} Rix, H.-W., \& Zaritsky, D.\ 1995,  
\apj,  447,  82 
 
\bibitem[Regan et al.(2001)]{regan01} Regan, M.~W., Thornley,
M.~D., Helfer, T.~T., Sheth, K., Wong, T., Vogel, S.~N., Blitz, L., \&
Bock, D.~C.-J.\ 2001, \apj, 561, 218

\bibitem[Rubin et al.(1999)]{rubin99} Rubin, V.~C., Waterman, A.~H., \&
Kenney, J.~D.~P. 1999, AJ 118, 236

\bibitem[Sage \& Wrobel(1989)]{sage89} Sage, L. J., \& Wrobel, J. M. 1989, \apj, 344, 204

\bibitem[Sault, Teuben, \& Wright(1995)]{sault95}Sault, R. J., Teuben, P. J., \& Wright, M. C. H. 1995, in ASP Conf.
     Ser. 77, Astronomical Data Analysis Software and Systems IV, ed. R.
     A. Shaw, H. E. Payne, \& J. J. E. Hayes (San Francisco: ASP), 433

\bibitem[Schoenmakers, Franx, \& de Zeeuw(1997)]{schoenmakers97}
Schoenmakers, R.~H.~M., Franx, M., \& de Zeeuw, P.~T. 1997, \mnras\ 292, 349

\bibitem[Somerville \& Primack(1999)]{somerville99} Somerville, R.~S., \& Primack, J.~R.\ 1999,  
\mnras,  310,  1087 
 

\bibitem[Swaters et al.(1999)]{swaters99} Swaters, R.~A., Schoenmakers,
R.~H.~M., Sancisi, R., \& van Albada, T.~S. 1999, \mnras\ 304, 330

\bibitem[Toomre(1964)]{Toomre64} Toomre, A. 1964. \apj 139, 1217

\bibitem[Tonry et al.(2001)]{tonry01} Tonry, J.~L., Dressler, A., Blakeslee, J.~P., Ajhar, E.~A., Fletcher, 
A.~B., Luppino, G.~A., Metzger, M.~R., \& Moore, C.~B.\ 2001,  
\apj,  546,  681 

\bibitem[Vollmer(2003)]{vollmer03} Vollmer, B. 2003, \aap, 398, 525

\bibitem[Warmels(1988)]{warmels1988} Warmels, R.~H. 1988, \aaps, 72, 57

\bibitem[Weiss et al.(2001)]{Weiss2001} Weiss, A., Ninninger, N.,
H\"uttemeister, S., \& Klein, U. 2001, \aap, 365, 571

\bibitem[Welch et al.(1996)]{welch96} Welch, W.~J.~et al.\
1996, \pasp, 108, 93

\bibitem[Wong(2001)]{wong01}Wong, T. 2001, Ph.D. thesis, University of California at Berkeley.

\bibitem[Zaritsky \& Rix(1997)]{zr97} Zaritsky, D., \& Rix, H.-W.\ 1997,  
\apj,  477,  118 
 
\end{thebibliography}
\end{document}